\documentclass[11pt,a4paper]{article}
\usepackage{jheppub}
\usepackage[T1]{fontenc}
\usepackage[utf8]{inputenc}
\usepackage{lmodern}
\usepackage{amsmath,amssymb,booktabs,array,microtype,etoolbox}
\usepackage{doi}
\usepackage{appendix}
\usepackage{listings}
\lstdefinestyle{databasejson}{
  basicstyle=\ttfamily\fontsize{10}{10}\linespread{1}\selectfont,
  columns=fullflexible,
  keepspaces=true,
  showstringspaces=false,
  upquote=true,
  breaklines=true,
  numbers=none,
  frame=none,
  xleftmargin=1em,
  aboveskip=.5\baselineskip,
  belowskip=.5\baselineskip
}
\AtBeginEnvironment{thebibliography}{\interlinepenalty=10000}
\newcommand{\A}{\mathcal A}
\newcommand{\CC}{\mathbb C}
\newcommand{\ZZ}{\mathbb Z}
\newcommand{\QQ}{\mathbb Q}
\newcommand{\Fix}{\operatorname{Fix}}
\newcommand{\OO}{\mathcal O}
\newcommand{\PP}{\mathbb P}
\toccontinuousfalse
\let\ReviewTableOfContents\tableofcontents
\renewcommand{\tableofcontents}{\begingroup\small\ReviewTableOfContents\endgroup}
\renewcommand{\afterTocSpace}{\medskip}
\renewcommand{\afterTocRuleSpace}{\medskip}
\hypersetup{pdftitle={Toric Calabi-Yau orientifolds: constructions, fixed loci and O-plane charges},pdfauthor={Xu Cao and Xin Gao},pdfsubject={Proceedings on toric Calabi-Yau orientifolds, divisor exchanges, reflections and database updates}}
\makeatletter
\gdef\@fpheader{Proceedings}
\patchcmd{\l@section}{1.0em}{0.2em}{}{\errmessage{TOC spacing patch failed}}
\makeatother
\title{Toric Calabi--Yau orientifolds:\\ constructions, fixed loci and O-plane charges}
\author{Xu Cao,}
\author{Xin Gao}
\affiliation{College of Physics, Sichuan University, Chengdu 610065, China}
\emailAdd{2026322020010@stu.scu.edu.cn}
\emailAdd{xingao@scu.edu.cn}
\abstract{
We review the systematic construction of  toric orientifold  Calabi-Yau threefolds through divisor exchanges and coordinate reflections. We relate the homogeneous-coordinate and lattice descriptions and explain the computation of fixed loci and the geometric data entering Hodge splittings and tadpole conditions. Our triangulation-wise analysis identifies multiple compatible involutions on a single triangulation and tracks their realisations across different triangulations, revealing the orientifold structure within and across toric phases. We complement this review with new calculations and explicit examples that refine the fixed-locus geometry and the associated orientifold charge data. The examples also identify coordinate representatives of the identity and invariant hypersurface families with unavoidable singularities. These results support an ongoing update of the orientifold catalogue and provide geometric input for D-brane configurations, flux compactifications and moduli stabilisation.
}

\begin{document}
\maketitle 

\section{Introduction}
\label{sec:intro}

 The existence of a unique Ricci-flat metric in each K\"ahler class of a compact K\"ahler manifold with vanishing first real Chern class~\cite{Calabi1957,Yau:1977ms,Yau:1978cfy} established a geometric foundation for supersymmetric string compactifications~\cite{Candelas:1985en}. Subsequent developments connected complex-structure periods to curve counting~\cite{Candelas:1990rm}, organised higher-genus topological-string amplitudes through holomorphic anomalies and Kodaira--Spencer theory~\cite{Bershadsky:1993ta,Bershadsky:1993cx}, and proposed a geometric interpretation of mirror symmetry through dual special Lagrangian torus fibrations~\cite{Strominger:1996it}. These connections motivate the systematic study of Calabi--Yau geometries together with the additional structures required for physical compactifications.

Orientifolds combine worldsheet parity with a target-space action~\cite{Sagnotti:1987tw,Brunner:2003zm}. For Type IIB compactifications, a suitable holomorphic involution of a Calabi--Yau threefold determines the surviving closed-string fields and constrains D-branes and their images~\cite{Grimm:2004uq,Jockers:2004yj}. In the O3/O7 sector, its isolated fixed points and fixed divisors support O3- and O7-planes. Their topology enters the tadpole conditions together with the D7 configuration and worldvolume flux~\cite{Blumenhagen:2008zz}. The resulting D3 tadpole constrains bulk flux stabilisation~\cite{Giddings:2001yu,Gukov:1999ya}, while divisor geometry enters nonperturbative effects and K\"ahler moduli stabilisation~\cite{Kachru:2003aw,Balasubramanian:2005zx}. These applications use geometric data for a specified involution and hypersurface family.

Toric hypersurfaces provide an explicit setting for obtaining these data. Batyrev's construction associates anticanonical Calabi--Yau hypersurface families to reflexive polytopes~\cite{Batyrev:1993oya}, while Cox's homogeneous-coordinate quotient makes their sections and coordinate identifications accessible to calculation~\cite{Cox:1993fz,CoxLittleSchenck}. The Kreuzer--Skarke classification~\cite{Kreuzer:2000xy}, resolved hypersurface databases and computational tools~\cite{Altman:2014bfa,Demirtas:2022hqf} support systematic searches. A fine, regular, star triangulation (FRST) specifies an ambient toric phase. Different triangulations can describe different K\"ahler chambers of the same Calabi--Yau geometry~\cite{Witten:1993yc}. An orientifold search must therefore keep track of both the action and the phase on which it is realised. Existing orientifold searches include divisor exchanges~\cite{Gao:2013pra,Altman:2021pyc}, single-coordinate reflections and their D3 charges ~\cite{Crino:2022zjk}, and exchanges together with multiple reflections~\cite{Cao:2024oqx}. Related constructions on complete intersections also include invariant conifold singularities and resolution branches~\cite{Carta:2020ohw}. A lattice-based description organises ambient involutions and their compatible toric phases~\cite{Moritz:2023jdb}.

We organise orientifold data by triangulation, recording the compatible divisor exchanges and the fixed-locus geometry of their invariant hypersurface families.  Building on the triangulation-wise constructions of~\cite{Altman:2021pyc,Cao:2024oqx}, we record both directions of this correspondence,
\begin{equation}
 T\longmapsto\mathcal S(T),
 \qquad
 \sigma\longmapsto\mathcal T(\sigma),
 \label{eq:phase-action-relation}
\end{equation}
where $\mathcal S(T)$ is the set of proper divisor exchanges supported by $T$, and $\mathcal T(\sigma)$ lists the supporting triangulations in the catalogue. In particular, we identify multiple distinct involutions supported by a single FRST, track a given involution across different triangulations, and determine the phases shared by different involutions. This provides a unified description of the many-to-many relation between toric phases and their compatible orientifold involutions.

This proceedings contribution combines a review of these construction methods with new checks and selected updates to the catalogue of~\cite{Cao:2024oqx}. Fixed points must be determined modulo the full group of homogeneous-coordinate identifications and then restricted to the invariant hypersurface. The examples show how incomplete identifications can miss fixed components, while deleting a fixed equation can retain points or branches that are not fixed. We obtain revised fixed-locus, Hodge-splitting and charge data, distinguish representatives of the identity from nontrivial involutions, and identify invariant families with unavoidable singularities. Sections~\ref{sec:construction} and~\ref{sec:examples} give the construction and worked examples. Sections~\ref{sec:catalogue} and~\ref{sec:discussion} describe the catalogue update and outlook. Appendix~\ref{app:database-format} gives the record format.

\section{Construction of toric Calabi--Yau orientifolds}
\label{sec:construction}\label{sec:review-geometry}\label{sec:geometry}

\subsection{Toric hypersurfaces and proper involutions}
\label{sec:review-hypersurfaces}\label{sec:review-actions}

Let $N\simeq\ZZ^4$ and $M=\operatorname{Hom}(N,\ZZ)$ be dual lattices, and let $\Delta^*\subset N_{\mathbb R}$ be a four-dimensional reflexive polytope with dual $\Delta\subset M_{\mathbb R}$. We take as primitive ray generators the boundary lattice points $n_1,\ldots,n_k$ that are not interior to facets, and write $\mathsf V=(n_1\ \cdots\ n_k)$. A fine, regular, star triangulation (FRST) of this point configuration together with the origin determines a complete simplicial fan $\Sigma$ and a projective toric fourfold $\A$. Fineness is understood relative to the selected point set.

The full group of homogeneous-coordinate identifications, which we call the Cox group, is $\mathcal G=\ker((\CC^*)^k\to T_N)$, where $T_N=\operatorname{Hom}(M,\CC^*)$ is the dense torus of $\A$ and the map is induced by $\mathsf V$. After choosing a splitting $\mathcal G\simeq(\CC^*)^r\times G$, Cox's geometric quotient is
\begin{equation}
 \A=\frac{\CC^k\setminus Z}{\mathcal G},
 \qquad
 x_i\sim f_i\left(\prod_{a=1}^{r}\lambda_a^{W_{ai}}\right)x_i,
 \qquad r=k-4,
 \label{eq:review-cox-quotient}
\end{equation}
where $x_i$ is the homogeneous coordinate associated with $n_i$ and $f=(f_1,\ldots,f_k)$ represents an element of the finite abelian factor $G$. The exponents $W_{ai}$ of the torus parameters $\lambda_a\in\CC^*$ form the gauged linear sigma model (GLSM) weight matrix $\mathbf W$. The embedded group $\mathcal G\subset(\CC^*)^k$ determines the coordinate identifications for any choice of splitting~\cite{Cox:1993fz,CoxLittleSchenck}.

The allowed coordinate set is the union, over maximal cones $\tau\in\Sigma$, of the open sets on which $x_i\neq0$ for every ray not contained in $\tau$. The Stanley--Reisner ideal $\mathcal I_{\rm SR}$ is generated by square-free monomials $x_{i_1}\cdots x_{i_s}$ corresponding to minimal sets of rays not contained in a common cone. Thus $ Z=\bigcup_{\{i_1,\ldots,i_s\}}
       \{x_{i_1}=\cdots=x_{i_s}=0\}.
$ The toric irrelevant ideal has zero set $Z$.

Both the continuous and finite identifications can be recovered from the Smith normal form $ U\mathsf V B=
 \bigl(\operatorname{diag}(d_1,\ldots,d_4)\ \ 0\bigr)$, with 
 $U\in GL(4,\ZZ), B\in GL(k,\ZZ)$ and $ d_j>0. $ The last $r$ columns of $B$ give an integral basis of $\ker\mathsf V$. Their transposes may be chosen as the rows of $\mathbf W$, so that $W_{ai}=B_{i,4+a}$ and $\mathbf W\mathsf V^{\mathsf T}=0$. For $j\leq4$, the one-parameter subgroup $x_i\mapsto t^{B_{ij}}x_i$ maps to the $j$th target-torus coordinate as $t^{d_j}$ in the transformed lattice basis. Its kernel supplies a finite generator whenever $d_j>1$:
\begin{equation}
 x_i\longmapsto e^{2\pi i B_{ij}/d_j}x_i.
 \label{eq:finite-identifications}
\end{equation}
Consequently, $G\simeq\prod_{j=1}^4\mu_{d_j}$, where $\mu_d$ is the group of $d$th roots of unity. Thus $G$ is trivial precisely when the rays generate $N$. This finite factor is distinct from the stabilizer of a coordinate representative, which can be nontrivial even when $G$ is trivial.

The anticanonical hypersurface $X=\{P=0\}\subset\A$ has equation
\begin{equation}
 P=\sum_{m\in\Delta\cap M}a_mM_m,
 \qquad
 M_m=\prod_{i=1}^{k}x_i^{\langle m,n_i\rangle+1}.
 \label{eq:review-anticanonical}
\end{equation}
The lattice-point expression incorporates both the GLSM weights and the finite grading of the anticanonical class. We write $\widehat D_i=\{x_i=0\}$ for the ambient coordinate divisors, $D_i=\widehat D_i|_X$ for their restrictions, and $H=-K_\A=\sum_i\widehat D_i$, so that $[X]=H$. We work with favourable hypersurfaces, whose $H^{1,1}(X)$ is spanned by the restricted toric divisor classes. For the selected ray set, $h^{1,1}(X)=k-4=r$~\cite{Batyrev:1993oya,Altman:2014bfa}. Divisor data used to select candidate exchanges are evaluated on a generic smooth member of the full anticanonical family.

Following~\cite{Cao:2024oqx}, nontrivial identical divisors (NIDs) have the same Hodge diamond and different divisor line bundles:
\begin{equation}
 h^{p,q}(D_i)=h^{p,q}(D_j)\quad\text{for all }p,q,
 \qquad
 \OO_X(D_i)\not\simeq\OO_X(D_j).
 \label{eq:NID}
\end{equation}

A candidate involution may exchange several disjoint NID pairs,
\[
 \sigma=\sigma_1\circ\cdots\circ\sigma_n,
 \qquad
 \sigma_m:x_{i_m}\longleftrightarrow x_{j_m}.
\]
We apply the compatibility conditions to the combined permutation. Let $\mathcal I_{\rm lin}$ be the ideal generated by the principal divisor relations $\sum_i\langle m,n_i\rangle\widehat D_i$, for $m\in M$, in $\QQ[\widehat D_1,\ldots,\widehat D_k]$. A proper NID involution satisfies
\begin{equation}
 \sigma(\mathcal I_{\rm SR})=\mathcal I_{\rm SR},
 \qquad
 \sigma(\mathcal I_{\rm lin})=\mathcal I_{\rm lin},
 \label{eq:properness}
\end{equation}
normalises the full group $\mathcal G$, and preserves the triple-intersection form on $X$. The normalisation condition ensures that the permutation descends through all coordinate identifications, including the finite ones. For an integral fan automorphism, intersection-form invariance follows from preservation of $H$. We retain it as a consistency check.

A coordinate reflection has the form $x_i\mapsto(-1)^{s_i}x_i$, with $s_i\in\{0,1\}$. It preserves the allowed coordinate set and commutes with $\mathcal G$, hence descends to a torus action on $\A$. It preserves every toric orbit and coordinate divisor setwise, without changing the fan. Distinct sign patterns may nevertheless define the same quotient action, and a nonempty sign pattern can represent the identity. These identifications are treated before counting reflection classes. On an invariant favourable hypersurface, reflections act trivially on the inherited divisor classes.

On a smooth invariant hypersurface $X$, we choose an invariant K\"ahler form, $\sigma^*J=J$. The holomorphic three-form then has parity $\sigma^*\Omega_3=\pm\Omega_3$, giving the Type IIB projections~\cite{Grimm:2004uq}
\begin{equation}
 \mathcal O=
 \begin{cases}
 (-1)^{F_L}\Omega_p\sigma,
 &\sigma^*\Omega_3=-\Omega_3\quad\text{(O3/O7)},\\
 \Omega_p\sigma,
 &\sigma^*\Omega_3=+\Omega_3\quad\text{(O5/O9)},
 \end{cases}
 \label{eq:review-orientifold}
\end{equation}
where $\Omega_p$ is worldsheet parity and $F_L$ is left-moving spacetime fermion number. At a fixed component of complex codimension $d$ in the smooth threefold, the differential of $\sigma$ is $+1$ along the component and $-1$ in its normal directions. Fixed surfaces and isolated points therefore support O7- and O3-planes, respectively, whereas fixed curves support O5-planes. The identity action on $X$ gives the O9 case. In each case the codimension is measured on $X$, after imposing the hypersurface equation.

\subsection{Invariant polynomials and fixed loci}
\label{sec:section-character}\label{sec:review-fixed-methods}\label{sec:restriction}

For a coordinate exchange or reflection, an invariant hypersurface is defined by a polynomial of definite parity:
\begin{equation}
P_{\rm sym}(\sigma x)=\gamma P_{\rm sym}(x),
\qquad \gamma=\pm1,
\label{eq:invariant-polynomial}
\end{equation}
where we take $\gamma=+1$ unless stated otherwise and choose generic coefficients within the specified eigenspace.

For a divisor exchange, let $\mathcal M$ be the monomial list used to construct $P_{\rm sym}$. Following~\cite{Cao:2024oqx}, we impose the following coefficient conditions for $\gamma=+1$:
\begin{equation}
 \begin{cases}
 a_m=0,&\sigma(M_m)\notin\mathcal M,\\
 a_m=a_n,&\sigma(M_m)=M_n,\quad n\ne m,\\
 a_m\ \text{arbitrary},&\sigma(M_m)=M_m.
 \end{cases}
 \label{eq:coefficient-regulation}
\end{equation}
For $\gamma=-1$, the last two lines become $a_m=-a_n$ and $a_m=0$, respectively. A proper exchange permutes the full anticanonical monomial set, so the first line is relevant only if $\mathcal M$ is restricted.

The same parity $\gamma$ enters the transformation of the holomorphic three-form. In the coordinate order of~\eqref{eq:review-cox-quotient}, write $\omega=dx_1\wedge\cdots\wedge dx_k$ and let $V^a$ generate the torus actions:
\begin{equation}
V^a=\sum_{i=1}^{k}W_{ai}x_i\frac{\partial}{\partial x_i},
\qquad a=1,\ldots,r.
\end{equation}
On a smooth invariant hypersurface the holomorphic three-form is the residue
\begin{equation}
\Omega_3=\frac{1}{2\pi i}
\oint_{P_{\rm sym}=0}
\frac{\omega\cdot\prod_{a=1}^{r}V^a}{P_{\rm sym}},
\label{eq:three-form}
\end{equation}
where the dot denotes successive contraction in the indicated order. For a coordinate exchange the numerator picks up the sign of the coordinate permutation times the determinant of the induced action on the span of the $V^a$. The denominator contributes $1/\gamma$. A reflection of $p$ coordinates fixes the $V^a$ and multiplies $\omega$ by $(-1)^p$, giving $\sigma^*\Omega_3=(-1)^p\Omega_3/\gamma$.

Before computing fixed loci, we identify reflections inducing the same quotient action. Two representatives $\sigma,\sigma'$ are equivalent precisely when a single $g=(\lambda,f)\in\mathcal G$ satisfies
\begin{equation}
 f_i\prod_{a=1}^{r}\lambda_a^{W_{ai}}
 =\frac{\sigma'(x_i)}{\sigma(x_i)},
 \qquad i=1,\ldots,k,
 \label{eq:reflection-class}
\end{equation}
where the ratios on the dense coordinate torus are constant signs. Taking $\sigma'$ to be the identity detects trivial quotient actions. The dense torus of $\A$ has four independent multiplicative coordinates. Their signs give at most $2^4=16$ classes, including the identity, among the $2^k$ sign choices on the homogeneous coordinates\footnote{
Writing a coordinate reflection as
$\sigma_s:x_i\mapsto(-1)^{s_i}x_i$, with
$s=(s_1,\ldots,s_k)^T\in\{0,1\}^k$, its induced torus action is
labelled by $c(s)=\mathrm V s\bmod 2N$.
Two sign patterns $s,s'$ define the same automorphism of $\mathcal A$
precisely when $\mathrm V(s-s')\in 2N$.
There are therefore $2^{r_2}-1$ nontrivial coordinate-reflection
classes, where
$r_2=\operatorname{rank}_{\mathbb F_2}(\mathrm V\bmod 2)$ and $\mathbb F_2=\mathbb Z/2\mathbb Z$.
These exhaust the fifteen nontrivial order-two elements of  $T_N[2]\simeq N/2N$. 
precisely when $r_2=4$.
For $r_2<4$, the remaining torus involutions require more general
coordinate phase representatives whose squares lie in the full
Cox group $\mathcal G$.
}.
 By~\eqref{eq:review-anticanonical}, equivalent representatives differing on $p$ coordinates multiply every anticanonical monomial by the same relative sign $(-1)^p$. Their polynomial eigenvalues must therefore be related by
\begin{equation}
 \gamma'=(-1)^p\gamma.
 \label{eq:reflection-eigenvalue}
\end{equation}
With this adjustment, the hypersurface family, fixed locus and three-form parity are unchanged.

An ambient point $[x]$ is fixed when $\sigma x=g\cdot x$ for some $g\in\mathcal G$. Its coordinate vanishing set must be allowed by $\mathcal I_{\rm SR}$ and preserved by the coordinate permutation. On the nonzero coordinates, the equations are
\begin{equation}
 f_i\prod_{a=1}^{r}\lambda_a^{W_{ai}}
 =\frac{\sigma(x_i)}{x_i},
 \qquad x_i\neq0,
 \quad f\in G,\quad \lambda_a\in\CC^*.
 \label{eq:reflection-fixed-weights}
\end{equation}
The compensating scaling can vary with the point. The right-hand sides are signs for reflections and coordinate ratios for exchanges. We solve over the complex torus for every $f\in G$ and group branches related by finite scalings on the quotient.

For exchanges with trivial $G$, we use the homogeneous parity polynomials from the Segre embedding of~\cite{Cao:2024oqx}. Besides unchanged coordinates and products of exchanged pairs, we use
\begin{equation}
 y_\pm(\boldsymbol a)
 =\prod_{\ell=1}^{n}x_{i_\ell}^{a_\ell}
 \pm\prod_{\ell=1}^{n}x_{j_\ell}^{a_\ell},
 \qquad
 \sum_{\ell=1}^{n}a_\ell
 (\mathbf w_{i_\ell}-\mathbf w_{j_\ell})=0,
 \label{eq:parity-generators}
\end{equation}
where $\mathbf w_i$ is column $i$ of $\mathbf W$, $\boldsymbol a\in\ZZ_{\geq0}^{n}\setminus\{0\}$, and the orientation of each exchanged pair may be chosen separately. The weight condition makes both terms homogeneous of the same degree. For a parity polynomial $y_\alpha$ of weight $\widetilde W_{a\alpha}$, a fixed point with $y_\alpha\neq0$ satisfies
\begin{equation}
 \prod_{a=1}^{r}\lambda_a^{\widetilde W_{a\alpha}}
 =\frac{\sigma(y_\alpha)}{y_\alpha}\in\{+1,-1\}.
 \label{eq:review-generator-compensation}
\end{equation}
We retain the algebraic relations among the auxiliary polynomials and verify candidates using~\eqref{eq:reflection-fixed-weights} in the original coordinates.

For a reduced irreducible ambient fixed component $\mathcal F$, its intersection with the hypersurface is
\begin{equation}
 \mathcal F_X=\mathcal F\cap X
 =\{P_{\rm sym}|_{\mathcal F}=0\}
 \subset\mathcal F.
 \label{eq:review-fixed-restriction}
\end{equation}
We use the reduced support of this intersection for geometric component and Euler-characteristic calculations. An identically vanishing restriction places all of $\mathcal F$ in $X$. A nonzero restriction has an empty or codimension-one zero locus. 
On each coordinate patch $U$ allowed by $\mathcal I_{\rm SR}$,
we solve the  compensation equations~\eqref{eq:reflection-fixed-weights}.
For each resulting fixed-locus branch, let
$\varphi_1(x)=\cdots=\varphi_p(x)=0$
denote its defining equations, including the coordinate
vanishings specifying the stratum.
Together with the hypersurface equation, they define
\begin{equation}
 \mathcal I_U
 =\langle P_{\rm sym},\varphi_1,\ldots,\varphi_p\rangle
 \subset R_U,
 \label{eq:fixed-ideal}
\end{equation}
where $R_U$ is the coordinate ring with the required nonzero coordinates inverted. A fixed equation $\varphi_j=0$ is redundant precisely when every solution of the remaining equations also satisfies it:
\begin{equation}
 P_{\rm sym}(x)=0,\quad
 \varphi_i(x)=0\ (i\ne j)
 \quad\Longrightarrow\quad
 \varphi_j(x)=0,\qquad x\in U.
\end{equation}
This condition is checked on each relevant patch. Section~\ref{sec:excess-curves} gives an explicit example where omitting a fixed equation introduces an additional component without changing the dimension.

On a pointwise fixed open curve, we count distinct nonzero roots in a coordinate on the quotient curve, after removing an invertible monomial factor. If a compact ambient fixed curve $\mathcal F$ is a reduced complete intersection of divisor classes $a,b,c$ and meets $X$ in finitely many transverse points in the smooth locus of $\A$, then
\begin{equation}
 N_{\mathcal F_X}
 =\int_{\A}abcH
 =\int_Xabc,
 \label{eq:curve-to-points}
\end{equation}
where $H=[X]$ and the classes on the right are restricted to $X$.

\subsection{Relation to lattice involutions}
\label{sec:integral-quotient}
\label{sec:slt-correspondence}

Fix the fan $\Sigma$ of an FRST for a favourable hypersurface. We compare individual proper NID exchanges with nonidentity lattice involutions $L$ that preserve $\Sigma$ and exchange rays only in pairs whose ambient divisors have distinct integral classes. We take the pure lattice action $L$, with trivial torus translation $[t]=0$ in the notation of~\cite{Moritz:2023jdb}. The corresponding actions on rays and coordinates are
\begin{equation}
L:n_{i_m}\longleftrightarrow n_{j_m},
\qquad
\sigma_m:x_{i_m}\longleftrightarrow x_{j_m},
\label{eq:lattice-exchange}
\end{equation}
where the coordinate involution $\sigma$ is the composition of the disjoint exchanges $\sigma_m$.

A permutation preserving the SR ideal descends to an automorphism of $\A$ precisely when it maps scaling orbits to scaling orbits, or equivalently when it normalises the full Cox group $\mathcal G$ in~\eqref{eq:review-cox-quotient}. Characters trivial on $\mathcal G$ are pulled back from the dense torus and have exponents in $\mathsf V^{\mathsf T}M\subset\ZZ^k$. A proper NID exchange  satisfies this condition by definition. Let $\mathsf V_\sigma$ be $\mathsf V$ with its columns permuted by $\sigma$. The rows of $\mathsf V$ and of $\mathsf V_\sigma$ give two integral bases of $\mathsf V^{\mathsf T}M$, so $\mathsf V_\sigma=L\mathsf V$ for a unique unimodular $L$. Since $\sigma^2=\mathrm{id}$ and the rays span $N_{\mathbb R}$, $L^2=\mathrm{id}$. Preservation of the SR ideal means that the ray permutation takes cones to cones, so $L$ preserves $\Sigma$. The exchanged ambient classes are distinct, since their restrictions satisfy the line-bundle condition in~\eqref{eq:NID}. Conversely, a lattice involution preserving $\Sigma$ induces a coordinate exchange preserving $\mathcal I_{\rm SR}$ and $\mathsf V^{\mathsf T}M$. It therefore normalises $\mathcal G$ and preserves $\mathcal I_{\rm lin}$, generated by the rational span of these principal divisor relations. It also permutes the $\widehat D_i$ and fixes their sum $H$, preserving the CY triple intersections $\int_\A\widehat D_i\widehat D_j\widehat D_\ell H$.

To check the NID conditions, the lattice action identifies a divisor on a generic anticanonical hypersurface with its partner on the image hypersurface. Their generic Hodge diamonds therefore agree. For the retained rays, restriction to a generic nondegenerate hypersurface introduces no additional integral divisor relations.\footnote{For the dense torus $T\subset\A$, $H^1(T,\ZZ)\simeq H^1(X\cap T,\ZZ)$~\cite{Batyrev:2005jc}. A rational function defining a principal relation among the $D_i$ is a unit on $X\cap T$, so its winding class comes from a character $m\in M$. Each retained ray has a positive-dimensional dual face, and a generic nondegenerate $X$ meets the corresponding open divisor orbit transversely. A meridian around this intersection represents $n_i$ in $H_1(T,\ZZ)$, giving order $\langle m,n_i\rangle$ along $D_i$. The relation is therefore already an ambient principal relation.} The distinct ambient classes consequently restrict to distinct line bundles on $X$, completing the converse.

The sign of the coordinate permutation is the product of its determinants on $\ker\mathsf V$ and on the quotient, where it acts by $L$. In~\eqref{eq:three-form}, the contraction with the GLSM vector fields gives a second factor of the determinant on $\ker\mathsf V$. This determinant squares to one, leaving $\sigma^*\Omega_3=(\det L)\Omega_3/\gamma$. Since the rays have convex hull $\Delta^*$, $L$ preserves $\Delta^*$ and $L^{\mathsf T}$ preserves the dual polytope $\Delta$. Substituting the exchanged coordinates sends $M_m$ to $M_{L^{\mathsf T}m}$ in~\eqref{eq:review-anticanonical}. The two constructions thus act on the same anticanonical monomials and give the same involution on $X$ for a given $P_{\rm sym}$.

The correspondence pairs individual pure exchanges with lattice involutions on a fixed ambient fan. For a specified $L$, the height construction of~\cite{Moritz:2023jdb} symmetrises the heights of a chosen FRST, producing an $L$-invariant fine, regular, star subdivision, which may itself be a triangulation. Our approach instead tests the NID candidates on each input FRST. Exhausting both the FRST list and the candidate exchanges gives the complete set of compatible pairs within this sector. Section~\ref{sec:involutions-across-triangulations} illustrates the resulting many-to-many relation.

\subsection{Smoothness, Euler characteristics and orientifold data}
\label{sec:topology}\label{sec:adjunction}\label{sec:orbit-euler}
\label{sec:lefschetz}\label{sec:tadpole}

Suppose every section in the chosen eigenspace vanishes on a smooth compact surface $S\subset\A$, with $\A$ smooth along $S$. Locally, write $S=\{u=v=0\}$ and $P_{\rm sym}=uA_1+vA_2$. The first normal derivative is a section of
\begin{equation}
 E=N^*_{S/\A}\otimes\OO_S(H),\qquad
 dP_{\rm sym}|_S=(A_1|_S,A_2|_S).
 \label{eq:normal-derivative}
\end{equation}
The zeros of this section are the singular points of the hypersurface on $S$. A nonzero $\int_S c_2(E)$ forces singularities in every member and counts them when the zeros are isolated and transverse~\cite{Moritz:2023jdb}. If $S$ is cut out regularly by two ambient divisors of classes $a,b$, then
\begin{equation}
 E=\OO_S(H-a)\oplus\OO_S(H-b),\qquad
 \int_Sc_2(E)=\int_{\A}ab(H-a)(H-b).
 \label{eq:normal-derivative-degree}
\end{equation}

We use $\chi$ for the topological Euler characteristic of the reduced underlying space. For a smooth fixed surface of class $s=[S]$ in a smooth Calabi--Yau threefold, adjunction gives
\begin{equation}
 \chi(S)=\int_Xs\bigl(c_2(X)+s^2\bigr).
 \label{eq:divisor}
\end{equation}
Here $s$ is the class of the reduced component. If the surface is a smooth regular intersection of two ambient divisors of classes $a,b$, with $\A$ smooth along it, then
\begin{equation}
 \chi(S)=\int_{\A}ab\left[c_2(\A)-c_1(\A)(a+b)+a^2+ab+b^2\right],
 \label{eq:ambient-surface}
\end{equation}
where $c_1(\A)=H$ and $c_2(\A)=\sum_{i<j}\widehat D_i\widehat D_j$ on the smooth ambient locus. This expression uses only ambient data along $S$.

Complete toric surfaces also admit a direct Euler calculation~\cite{CoxLittleSchenck}. For the surface $S=V(\tau)$ associated with a two-dimensional cone $\tau$,
\begin{equation}
 \chi(S)=\#\{\text{maximal cones of }\Sigma\text{ containing }\tau\}.
 \label{eq:toric-surface-euler}
\end{equation}
Each zero-dimensional orbit contributes one, while positive-dimensional torus orbits have vanishing compactly supported Euler characteristic. The formula remains valid for singular toric surfaces and counts coarse geometric points without stabilizer weights.

For the Hodge and charge formulas below, let $X$ be a smooth connected compact Calabi--Yau threefold with $h^{0,1}(X)=h^{0,2}(X)=0$ and holomorphic involution $\sigma$. Define
\begin{equation}
  H^{p,q}_{\pm}(X)
=
\ker\!\left(\sigma^*\mp\mathrm{id}\right)
\subset H^{p,q}(X),
\qquad
h^{p,q}_{\pm}=\dim_{\mathbb C}H^{p,q}_{\pm}(X).  
\end{equation}
In the favourable setting, the odd part of $H^{1,1}(X)$ is spanned by the nonzero differences of exchanged divisor classes, after imposing their linear relations. Coordinate reflections act trivially on the inherited classes. The Hodge decomposition and the topological Lefschetz relation give
\begin{equation}
h^{2,1}_{-}
=
h^{1,1}_{-}
+
\frac{\chi(\Fix\sigma)-\chi(X)}{4}
-
h^{3,0}_{-},
\label{eq:lefschetz}
\end{equation}
where $\chi(X)=2(h^{1,1}-h^{2,1})$ and the fixed locus is taken on $X$. The odd dimension $h^{3,0}_-$ is $1$ in the O3/O7 sector and $0$ in the O5/O9 sector.

For a smooth O3/O7 hypersurface, let $N_{\rm O3}$ denote the number of isolated fixed points. The holomorphic Lefschetz formula relates this number to the fixed divisors $S$~\cite{AtiyahBott1968,Atiyah:1968rj,Collinucci:2008pf}:
\begin{equation}
3N_{\rm O3}
+\sum_S\left(
\chi(S)-2\int_X[S]^3
\right)=48.
\label{eq:index48}
\end{equation}
For O3$^-$- and O7$^-$-planes, the D7 homology condition is
\begin{equation}
 \sum_aN_a([D_a]+[D'_a])=8\sum_S[S],
 \label{eq:d7-tadpole}
\end{equation}
where $N_a$ counts the D7-branes on $D_a$ and $D'_a=\sigma(D_a)$. We specialise to four D7-branes and their four images on each O7 surface, giving eight D7 sheets in the double cover and cancelling the D7 charge locally. For this configuration, the D3 tadpole condition is
\begin{equation}
 N_{\rm D3}+\frac12N_{\rm flux}+N_{\rm gauge}
 =\frac{N_{\rm O3}+\sum_S\chi(S)}4
 \equiv L_{\rm D3}.
 \label{eq:budget}
\end{equation}
Here $N_{\rm D3}$ and the D7 gauge-flux contribution $N_{\rm gauge}$ are measured in quotient units, whereas the three-form flux charge $N_{\rm flux}$ is evaluated on the double cover. The right-hand side is the geometric tadpole from O3-planes and the curvature charges of the smooth O7 and D7 worldvolumes~\cite{Green:1996dd,Morales:1998ux}.

For a given triangulation, the fixed-locus calculation proceeds as follows:
\begin{enumerate}
 \setlength{\itemsep}{0pt}
 \item Retain NID exchanges that preserve the SR and linear divisor ideals and the triple-intersection form on $X$, and normalise the full group of toric scalings. For coordinate reflections, retain one representative of each nontrivial equivalence class. In each nonzero parity sector under consideration, choose a general $P_{\rm sym}$.
 \item Solve the fixed equations modulo the full group of toric scalings. Impose $P_{\rm sym}=0$ together with the fixed equations on each allowed stratum to obtain the components on $X$.
 \item Decompose the reduced fixed locus, count isolated points and
 compute the Euler characteristics of its components.
\end{enumerate}

\section{Concrete examples}\label{sec:examples}\label{sec:fixed}

We revisit two examples from~\cite{Cao:2024oqx} and use further cases to show how toric scalings and the restriction to $X$ affect the fixed loci. The reflection examples also show how singularities can persist throughout an invariant family. The final example examines how the set of proper divisor exchanges changes with the triangulation. We identify each geometry by $(h^{1,1},\text{polytope ID},\text{triangulation ID})$. The geometries of Examples A and B use the published coordinate order $x_1,\ldots,x_k$. The others use the database order $x_0,\ldots,x_{k-1}$.

\subsection{Example A: divisor exchange, fixed points and D3 charge}
\label{sec:examplea}\label{sec:charge-examples}

The geometry $(7,545,5)$ has Hodge numbers $(7,37)$. We revisit the divisor exchange of Example A in~\cite{Cao:2024oqx}. A general invariant hypersurface meets an invariant ambient curve in three points. The involution fixes one of the points and exchanges the other two.

\paragraph{Toric data and invariant family.}
In the order $x_1,\ldots,x_{11}$, the rays and weights are
\begin{equation}
 \mathsf V=\begin{pmatrix}
 1&1&1&1&-2&1&1&1&1&1&1\\
 0&2&0&0&-1&1&0&0&1&1&0\\
 0&0&2&0&-1&0&1&0&1&0&1\\
 0&0&0&2&-1&0&0&1&0&1&1
 \end{pmatrix},
 \qquad
 \mathbf W=\begin{pmatrix}
 1&0&1&0&2&0&0&0&0&2&0\\
 0&1&1&0&0&0&0&0&-2&0&0\\
 0&0&-1&1&0&0&0&0&2&-2&0\\
 0&0&1&0&1&1&0&0&-1&1&0\\
 0&0&0&0&1&0&1&0&0&1&0\\
 0&0&0&0&1&0&0&1&1&0&0\\
 0&0&-1&0&0&0&0&0&1&-1&1
 \end{pmatrix},
 \label{eq:exampleA-matrices}
\end{equation}
The anticanonical multidegree is $(6,0,0,3,3,3,0)$. The triangulation is specified by
\begin{equation}
\begin{aligned}
 \mathcal I_{\rm SR}=\langle\,
 &x_1x_2,x_1x_3,x_1x_4,x_1x_9,x_1x_{10},x_1x_{11}, x_2x_3,x_2x_4,x_2x_7,x_2x_8,x_2x_{11},\\
 & x_3x_4,x_3x_6,x_3x_8,x_3x_{10}, x_6x_{11},x_7x_{10},x_8x_9,x_8x_{10},x_8x_{11},x_{10}x_{11},\\
 &x_4x_5x_6x_7,x_4x_5x_6x_9,x_4x_5x_7x_9,
 x_5x_6x_7x_8,x_5x_6x_9x_{10},x_5x_7x_9x_{11}\,\rangle.
\end{aligned}
\label{eq:exampleA-SR}
\end{equation}
We consider the proper NID involution that exchanges three pairs simultaneously:
\begin{equation}
\sigma:\quad
x_1\longleftrightarrow x_2,\quad
x_7\longleftrightarrow x_9,\quad
x_8\longleftrightarrow x_{10}.
\label{eq:exampleA-involution}
\end{equation}
Exchanging $x_1$ and $x_2$ alone sends the SR generator $x_1x_9$ to $x_2x_9$, which is not in the ideal. The combined exchange sends it to $x_2x_7$ and preserves the SR and linear ideals and the triple-intersection form. The 40 anticanonical monomials give 28 even orbit sums by~\eqref{eq:coefficient-regulation}. The weight matrix $\mathbf W$ in~\eqref{eq:exampleA-matrices} gives determinant $+1$ for the induced action on the GLSM vector fields. The numerator of~\eqref{eq:three-form} therefore picks up only the sign of the odd coordinate permutation, giving $\sigma^*\Omega_3=-\Omega_3$ for the even polynomial.

\paragraph{Fixed equations and restriction to $X$.}
The rays $n_1,n_6,n_7,n_8$ in~\eqref{eq:exampleA-matrices} form a lattice basis of $N$, so $G$ is trivial and we use the homogeneous parity generators of~\eqref{eq:parity-generators}. For $x_1^m x_7^n x_8^p\pm x_2^m x_9^n x_{10}^p$, the weight condition reduces to $m=2n=2p$, with smallest nonzero solution $(m,n,p)=(2,1,1)$. The two exchanged monomials have common multidegree $(2,0,0,0,1,1,0)$. The generators are
\begin{equation}
\begin{gathered}
y_1=x_8x_{10},\quad
y_2=x_7x_9,\quad
y_3=x_3,\quad
y_4=x_4,\quad
y_5=x_5,\quad
y_6=x_6,
\\
y_7=x_1x_2,\quad
y_8=x_1^2x_7x_8+x_2^2x_9x_{10},\quad
y_9=x_1^2x_7x_8-x_2^2x_9x_{10},\quad
y_{10}=x_{11}.
\end{gathered}
\label{eq:exampleA-generators}
\end{equation}
Only $y_9$ is odd. Setting $y_9=0$ makes the two monomials in~\eqref{eq:exampleA-generators} equal. On this branch the fixed equations give the surface $\mathcal F_1=\{y_9=0\}\subset X$. The other branches give the isolated fixed sets in table~\ref{tab:exampleA}. 

The subset $\mathcal F_2$ lies on the invariant ambient curve $C=\{x_4=x_5=x_{11}=0\}$ and is selected by the additional equation $y_8=0$. On its open stratum, the homogeneous ratio
\begin{equation}
t=\frac{x_2^2x_9x_{10}}{x_1^2x_7x_8}
\label{eq:exampleA-curve-coordinate}
\end{equation}
is a coordinate after the torus identifications. The torus actions allow the choice
\begin{equation}
x_1=x_2=x_3=x_6=x_7=x_8=x_9=1,
\qquad x_{10}=t\ne0.
\label{eq:F2-coordinate-slice}
\end{equation}
The seven weight columns used here have determinant $\pm1$, so different values of $t$ represent different points on this open curve. The exchange acts as $t\mapsto t^{-1}$. The fixed equations give the torus parameters $\lambda=(\lambda_1,\ldots,\lambda_7)$ as
\begin{equation}
\lambda=(1,1,t^{-1},1,1,t,t),
\qquad t=t^{-1}.
\label{eq:F2-lambda}
\end{equation}
Thus the ambient fixed points on this open curve are $t=\pm1$. The point $t=1$ lies on $y_9=0$, whereas $y_8=1+t=0$ selects $t=-1$.

Restricting the invariant polynomial to the same curve gives
\begin{equation}
 P_{\rm sym}|_C=A(1+t^3)+B(t+t^2)
 =(1+t)\left[At^2+(B-A)t+A\right],
 \label{eq:f2-family}
\end{equation}
after removing an invertible monomial. The four terms arise from two exchange orbits, with independent coefficients $A,B$. For $A\ne0$ and $B\notin\{-A,3A\}$, there are three simple roots. The root $t=-1$ gives $\mathcal F_2$, and $\sigma$ exchanges the two roots of the quadratic factor.

The distinction is also visible in the intersection degrees:
\begin{equation}
 \int_{\A}\widehat D_4\widehat D_5\widehat D_{11}
       (2\widehat D_1+\widehat D_7+\widehat D_8)=1,
 \qquad
 \int_{\A}\widehat D_4\widehat D_5\widehat D_{11}H=3.
 \label{eq:f2-degrees}
\end{equation}
The first degree counts the fixed root $t=-1$. The second counts all three points of $C\cap X$. In the original calculation~\cite{Cao:2024oqx}, dropping $y_8=0$ from this branch retained the two points exchanged by $\sigma$.

The equations $x_3=x_5=x_{11}=y_8=0$ defining $\mathcal F_3$ give the same restriction and contribute one point. The equations $x_4=x_6=y_8=0$ instead define an ambient fixed curve. On a smooth coordinate chart, its hypersurface restriction is a cubic in the remaining coordinate $x_5$, with four independent coefficients. For general coefficients it has three distinct nonzero roots, giving the three points of $\mathcal F_4$:
\begin{equation}
 N_{\mathcal F_4}=
 \int_{\A}\widehat D_4\widehat D_6
             (2\widehat D_1+\widehat D_7+\widehat D_8)H=3.
 \label{eq:f4-degrees}
\end{equation}

\begin{table}[tb]
\centering\small
\begin{tabular}{@{}lll@{}}
\toprule
Fixed set & Equations in $X$ & Contribution to $4L_{\rm D3}$\\
\midrule
$\mathcal F_1$ & $y_9=0$ & O7 surface, $\chi=27$\\
$\mathcal F_2$ & $x_4=x_5=x_{11}=y_8=0$ & 1 O3 point\\
$\mathcal F_3$ & $x_3=x_5=x_{11}=y_8=0$ & 1 O3 point\\
$\mathcal F_4$ & $x_4=x_6=y_8=0$ & 3 O3 points\\
\bottomrule
\end{tabular}
\caption{Fixed loci of the divisor exchange in
Example A, with $P_{\rm sym}=0$ imposed throughout.
For the local D7 configuration,
$4L_{\rm D3}=N_{\rm O3}+\sum_S\chi(S)=32$.}
\label{tab:exampleA}
\end{table}

The orbit sections have common zero set $x_5=x_{11}=y_8=0$. A general member avoids the unique ambient quotient singularity, where the retained monomial $x_5^3$ is nonzero. Away from the base curve the system is base-point-free, so a general member is smooth there. Along the curve, the SR conditions and $y_8=0$ make both monomials in $y_8$ nonzero. The ratio $t$ is transverse to the base curve on $x_5=x_{11}=0$, where the restriction takes the form~\eqref{eq:f2-family} up to an invertible factor. Its derivative at $t=-1$ is $3A-B$, nonzero for general coefficients, so the general member is also smooth along the curve.

\paragraph{Topological data and D3 tadpole.}
The equation $y_9=0$ cuts out $\mathcal F_1$ on $X$ with multiplicity one. Its class and intersection numbers are
\begin{equation}
s=2D_1+D_7+D_8,\qquad \int_Xs^3=-3,\qquad \int_Xc_2(X)s=30.
 \label{eq:exampleA-surface}
\end{equation}
Eq.~\eqref{eq:divisor} gives $\chi(\mathcal F_1)=30-3=27$. The remaining fixed sets contribute $N_{\rm O3}=1+1+3=5$ isolated points. The components are disjoint, so
\begin{equation}
\chi(\Fix\sigma)
=\chi(\mathcal F_1)+N_{\rm O3}
=27+5=32.
\label{eq:exampleA-fixed-euler}
\end{equation}
The holomorphic Lefschetz relation \eqref{eq:index48} is also satisfied:
\begin{equation}
3N_{\rm O3}
+\chi(\mathcal F_1)-2\int_X s^3
=3\cdot5+27-2(-3)=48.
\label{eq:exampleA-index}
\end{equation}
The local D3 tadpole follows from~\eqref{eq:budget}:
\begin{equation}
L_{\rm D3}
=\frac{\chi(\Fix\sigma)}4
=\frac{32}{4}=8.
\label{eq:exampleA-budget}
\end{equation}
The updated entries are therefore $N_{\rm O3}=5$ and $L_{\rm D3}=8$, replacing the previously tabulated values $N_{\rm O3}=9$ and $L_{\rm D3}=9$ in~\cite{Cao:2024oqx}.

The differences of the exchanged divisor classes satisfy
\begin{equation}
 2(D_1-D_2)+(D_7-D_9)+(D_8-D_{10})=0.
\end{equation}
There are two independent odd classes, so $(h^{1,1}_+,h^{1,1}_-)=(5,2)$. With $\chi(X)=2(7-37)=-60$ and $\sigma^*\Omega_3=-\Omega_3$, ~\eqref{eq:lefschetz} yields
\begin{equation}
 h^{2,1}_-=2+\frac{32-(-60)}{4}-1=24,\qquad
 (h^{2,1}_+,h^{2,1}_-)=(13,24).
 \label{eq:exampleA-hodge}
\end{equation}

Table~\ref{tab:identity-reflections} lists the three reflections reported as freely acting in Example A of~\cite{Cao:2024oqx}. In each case a single Cox scaling reproduces the action on all eleven coordinates. The induced action on $\A$ is therefore the identity and fixes every point of $X$.
\begin{table}[tb]
\centering\small
\begin{tabular}{@{}lll@{}}
\toprule
Reflected coordinates & Torus parameters $\lambda$ & $\gamma$\\
\midrule
$\{x_1,x_2\}$ & $(-1,-1,1,1,1,1,1)$ & $+1$\\
$\{x_1,x_3\}$ & $(-1,1,1,1,1,1,1)$ & $+1$\\
$\{x_5,x_8,x_9\}$ & $(1,1,1,1,1,-1,1)$ & $-1$\\
\bottomrule
\end{tabular}
\caption{Three coordinate representatives of the identity for the
weight matrix $\mathbf W$ in~\eqref{eq:exampleA-matrices}. The
polynomial eigenvalue is determined by its multidegree.}
\label{tab:identity-reflections}
\end{table}
In the last row, the anticanonical polynomial has eigenvalue $-1$ because the sixth component of its multidegree is odd. The numerator in~\eqref{eq:three-form} acquires the same sign, so $\Omega_3$ is unchanged.

\subsection{Example B: complex torus actions at a singular fixed point}
\label{sec:exampleB}

The reflection $\sigma:x_6\mapsto-x_6$ on the geometry $(3,79,0)$ has an isolated fixed point $\mathcal F=\{x_1=x_2=x_3=x_5=0\}$ that requires complex torus parameters to identify~\cite{Cao:2024oqx}. Every even invariant hypersurface passes through this point and is singular there. In the published order $x_1,\ldots,x_7$, the toric data are
\begin{equation}
 \mathbf W=\begin{pmatrix}
 1&0&1&1&0&0&0\\
 0&1&0&1&1&2&0\\
 0&0&0&0&0&1&1
 \end{pmatrix},\qquad\deg P=(3,5,2),
 \label{eq:exampleB-weights}
\end{equation}
\begin{equation}
 \mathcal I_{\rm SR}
 =\langle x_1x_3x_4,\ x_1x_3x_7,\ x_2x_4x_5,
              \ x_2x_5x_6,\ x_6x_7\rangle,
 \label{eq:exampleB-SR}
\end{equation}
where $G$ is trivial, and $P_{\rm sym}$ contains the monomials with even $x_6$ exponent.

At $\mathcal F$ the SR ideal forces $x_4,x_6,x_7\ne0$. The fixed equations~\eqref{eq:reflection-fixed-weights} reduce to
\begin{equation}
 \lambda_1\lambda_2=1,\qquad
 \lambda_2^2\lambda_3=-1,\qquad\lambda_3=1.
 \label{eq:exampleB-lambda}
\end{equation}
The last two equations give $\lambda_2^2=-1$, so a search over real torus parameters misses this point. The two solutions $\lambda=(-i,i,1)$ and $(i,-i,1)$ correspond to the same fixed point $\mathcal F$. Setting $x_4=x_6=x_7=1$ leaves a residual scaling $\lambda=(-1,-1,1)$, which reverses the four local coordinates $(u_1,u_2,u_3,u_4)=(x_1,x_2,x_3,x_5)$. The ambient chart is therefore $\CC^4/\ZZ_2$.

A monomial surviving at $\mathcal F$ uses only $x_4,x_6,x_7$. The multidegree of the hypersurface $(3,5,2)$ permits only the monomial $x_4^3x_6x_7$, which is odd under $\sigma$. Therefore its coefficient  vanishes in $P_{\rm sym}$, so every invariant member passes through $\mathcal F$. The multidegree $(3,5,2)$ makes the local polynomial invariant under the residual action $u\mapsto-u$, since its character is  $(-1)^3(-1)^5=1$. Its expansion therefore begins at quadratic order in $u$.

The quotient is described by the ten invariant coordinates $v_{ij}=u_i u_j$, $1\leq i\leq j\leq4$, whose defining relations are quadratic. The hypersurface equation takes the form
\begin{equation}
 P_{\rm loc}=\sum_{i\leq j}c_{ij}v_{ij}+O(v^2),
 \label{eq:exampleB-local-polynomial}
\end{equation}
and imposes at most one linear relation among these coordinates. Consequently,
\begin{equation}
 \dim_{\CC}T_{\mathcal F}X\geq\binom{5}{2}-1=9>3.
 \label{eq:exampleB-tangent}
\end{equation}
This proves that $\mathcal F$ is singular on every even member and updates the interpretation of the point included in the original O3 count.

\subsection{Eight isolated fixed points in a finite toric quotient}
\label{sec:finite-example}

A finite coordinate scaling can compensate a reflection even when the connected torus cannot. In the geometry $(4,4,0)$, the connected torus accounts for the two fixed divisors in the catalogue of~\cite{Cao:2024oqx}. Including $G$ in the fixed equations gives additional branches, which contribute eight isolated O3-planes on $X$. 

\paragraph{Finite action and invariant sections.} 
The rays and weights in the order $x_0,\ldots,x_7$ are
\begin{equation}
\begin{split}
 \mathsf V&=\begin{pmatrix}
 -2&-1&0&0&1&1&-1&0\\
 0&-3&0&1&0&1&0&-1\\
 -1&-1&1&0&0&1&0&0\\
 0&-2&0&0&0&2&0&0
 \end{pmatrix},\\
 \mathbf W&=\begin{pmatrix}
 1&0&1&0&2&0&0&0\\
 0&1&0&2&0&1&0&0\\
 0&0&0&0&1&0&1&0\\
 0&0&0&1&0&0&0&1
 \end{pmatrix},\qquad\deg P=(4,4,2,2),
\end{split}
\label{eq:finite-geometry}
\end{equation}
with
\begin{equation}
 \mathcal I_{\rm SR}
 =\langle x_0x_2,\ x_1x_5,\ x_3x_7,\ x_4x_6\rangle.
 \label{eq:finite-SR}
\end{equation}
The rays have even fourth component, and $\det(n_0,n_1,n_3,n_4)=-2$. The ray sublattice therefore has index two, so $G\simeq\ZZ_2$. The nonzero coset is represented by $(n_0+n_1+n_3+n_4)/2=(-1,-1,-1,-1)^{\mathsf T}$. Exponentiating the four coefficients $1/2$ gives the minus signs in the generator
\begin{equation}
 f=(-1,-1,1,-1,-1,1,1,1).
 \label{eq:finite-generator}
\end{equation}
The weights allow 81 monomials of degree $(4,4,2,2)$. Since $\prod_i f_i=+1$, the anticanonical monomials on $\A$ are the 41 with even total exponent on $x_0,x_1,x_3,x_4$. For the reflection $\sigma:x_0\mapsto-x_0$, the 28 even in $x_0$ give $P_{\rm sym}$.

\paragraph{Fixed curves and quotient roots. }
On the open stratum with $x_1=x_3=x_4=0$ and all remaining coordinates nonzero, the columns of $\mathbf W$ for $x_0$ and $x_2$ are equal. A torus action alone would require both $\lambda_1=-1$ and $\lambda_1=1$. The finite coordinate scaling~\eqref{eq:finite-generator} supplies the minus sign on $x_0$ and leaves the other nonzero coordinates unchanged, so~\eqref{eq:reflection-fixed-weights} holds with $\lambda_a=1$.

The torus actions set $x_2=x_5=x_6=x_7=1$, leaving $x_0\ne0$. The generator of $G$ then identifies $x_0$ with $-x_0$, so $z=x_0^2$ labels distinct quotient points. The same calculation gives quadratic restrictions on the four curves in table~\ref{tab:finite-cox-points}, which displays one explicit coefficient specialisation. The second and fourth rows retain the same three anticanonical monomials, so their restrictions coincide.

\begin{table}[tb]
\centering\small
\begin{tabular}{@{}lrrr@{}}
\toprule
Vanishing coordinates & Restricted polynomial & Discriminant & Points\\
\midrule
$\{x_1,x_3,x_4\}$ & $98z^2+93z+63$ & $-16047$ & 2\\
$\{x_1,x_4,x_7\}$ & $51z^2+18z+8$ & $-1308$ & 2\\
$\{x_3,x_4,x_5\}$ & $93z^2+39z+71$ & $-24891$ & 2\\
$\{x_4,x_5,x_7\}$ & $51z^2+18z+8$ & $-1308$ & 2\\
\bottomrule
\end{tabular}
\caption{The four fixed curves whose hypersurface intersections give isolated points in section~\ref{sec:finite-example}. In each row $z$ is a coordinate on the curve after all identifications.}
\label{tab:finite-cox-points}
\end{table}
Each quadratic has two distinct nonzero roots for general coefficients, as shown by the specialisation in table~\ref{tab:finite-cox-points}. The representatives $x_0=\pm\sqrt z$ define the same quotient point, so the four curves contribute eight isolated fixed points. The remaining four ambient fixed curves outside the fixed divisors $x_0=0$ and $x_2=0$ have $x_6=0$ and invertible monomial restrictions, so they do not meet a general hypersurface. The retained monomials have no common zero, so a general invariant hypersurface is smooth on the smooth locus of $\A$. The ambient singularities are isolated torus fixed points, at each of which a retained monomial is nonzero. A general member avoids these points and is therefore smooth everywhere. Since $\sigma^*\Omega_3=-\Omega_3$ by~\eqref{eq:three-form}, the eight isolated fixed points support O3-planes.

\subsection{Beyond dimension counting: an extraneous fixed-locus branch}
\label{sec:excess-curves}

The fixed-locus calculation in~\cite{Cao:2024oqx} used the dimension of the solution set to identify redundant fixed-point equations. The following example shows that dimension counting alone does not determine the full fixed locus. Omitting one equation introduces an additional curve of the same dimension. We illustrate this using the geometry $(2,1,0)$, whose GLSM weight matrix and SR ideal, in the coordinate order $x_0,\ldots,x_5$, are
\begin{equation}
 \begin{gathered}
 \mathbf W=\begin{pmatrix}1&0&0&0&1&1\\0&1&1&1&0&0\end{pmatrix},
 \qquad\deg P=(3,3),\\[2pt]
 \mathcal I_{\rm SR}=\langle x_0x_4x_5,\ x_1x_2x_3\rangle.
 \end{gathered}
 \label{eq:curve-example-geometry}
\end{equation}
The finite group $G\simeq\ZZ_3$ of coordinate scalings is generated by
\begin{equation}
f=(1,1,\zeta,\zeta^2,\zeta^2,\zeta),\qquad
\zeta=e^{2\pi i/3},
 \label{eq:curve-finite-action}
\end{equation}
so the ambient space is a finite quotient of $\PP^2\times\PP^2$. We take a general bicubic invariant under the proper exchange
\begin{equation}
\sigma:\quad
x_0\longleftrightarrow x_1,\quad
x_4\longleftrightarrow x_2,\quad
x_5\longleftrightarrow x_3.
\label{eq:curve-example-action}
\end{equation}
The odd coordinate permutation changes the sign of $\omega$. The exchange $V^1\leftrightarrow V^2$ gives a second minus sign in the numerator of~\eqref{eq:three-form}, so $\sigma^*\Omega_3=\Omega_3$ for the even bicubic. The symmetrised products of coordinate cubes have no common zero on the cover. A general bicubic is therefore smooth there and avoids the nine fixed points of $G$. Its quotient is smooth and the fixed curves support O5-planes.

 On the cover $\PP^2\times\PP^2$, three homogeneous equations for the diagonal fixed set are
\begin{equation}
 E_1=x_3x_4-x_2x_5,\qquad
 E_2=x_0x_3-x_1x_5,\qquad
 E_3=x_0x_2-x_1x_4.
 \label{eq:excess-ideal}
\end{equation}
They are related by
\begin{equation}
 x_3E_3=x_2E_2-x_1E_1,\qquad
 x_5E_3=x_4E_2-x_0E_1.
 \label{eq:curve-relations}
\end{equation}
The first two equations imply the third wherever $x_3$ or $x_5$ is nonzero. At $x_3=x_5=0$, both vanish without imposing $E_3$. The common zero set of $E_1$ and $E_2$ on the cover is thus the union of the diagonal and the surface $B_0=\{x_3=x_5=0\}$. To obtain the fixed set on $X$, we take the union of the three $G$-translates of $E_1=E_2=E_3=0$, intersect with the hypersurface and descend to the quotient.

The ambient fixed locus lifts to the three graphs
\begin{equation}
[x_1:x_2:x_3]
=
[x_0:\zeta^{2j}x_4:\zeta^j x_5],
\qquad j=0,1,2.
\label{eq:bicubic-fixed-graphs}
\end{equation}
Each meets the bicubic in a plane sextic, and $f$ cyclically permutes these three curves. The graphs meet at the three matching coordinate vertices of the two projective planes. The invariant bicubic contains a pure-cube product nonzero at each vertex, so a general member avoids these intersections. The sextics descend to a single smooth curve $T$ of genus ten, with
\begin{equation}
 \chi(T)=2-2\cdot10=-18.
\end{equation}

The extra surface $B_0\simeq\PP^1\times\PP^1$ instead cuts out a $(3,3)$ curve of genus four and Euler characteristic $-6$. The $\ZZ_3$ action is free on this curve for general coefficients, so its image $B$ has $\chi(B)=-2$. The three sextics meet the $(3,3)$ curve in eighteen points on the cover, giving six quotient points in $T\cap B$. Thus the two-equation locus would have
\begin{equation}
 \chi(T\cup B)=\chi(T)+\chi(B)-\#(T\cap B)
 =-18-2-6=-26.
 \label{eq:excess-euler}
\end{equation}
Using all three diagonal equations before taking the $G$-translates leaves only the fixed curve $T$ on the quotient. The extra component changes the Euler characteristic while leaving the dimension unchanged.

\subsection{Fixed surfaces and forced singularities}
\label{sec:singular-surfaces}

The next two reflections force smooth ambient surfaces into every even hypersurface. The bundles $E$ of~\eqref{eq:normal-derivative} have nonzero second Chern numbers, so every member of either family is singular. The ambient geometry determines the Euler characteristics of the reduced fixed surfaces.

\paragraph{A fixed $\PP^2$ in the geometry of Example A.}
Using the data of section~\ref{sec:examplea}, take
\begin{equation}
 \sigma:(x_3,x_8,x_9)\longmapsto(-x_3,-x_8,-x_9).
 \label{eq:exampleA-reflection}
\end{equation}
A monomial surviving on $S=\{x_1=x_5=0\}$ has $e_1=e_5=0$. The first and sixth rows of $\mathbf W$ in~\eqref{eq:exampleA-matrices} give $e_3+2e_{10}=6$ and $e_8+e_9=3$, so every such monomial is odd. The even sections therefore have the form
\begin{equation}
 P_{\rm sym}=x_1P_1+x_5P_5.
 \label{eq:exampleA-reflection-polynomial}
\end{equation}
The compensator $\lambda=(-1,1,1,1,1,-1,1)$ reproduces the reflection on $S$, making it pointwise fixed in every member.

The SR data identify $S\simeq\PP^2$ in the smooth ambient locus. Writing $h$ for its hyperplane class, the restrictions are $\widehat D_1|_S=-h$, $\widehat D_5|_S=h$ and $H|_S=3h$. Surface adjunction gives $\chi(S)=3$, while the normal derivative is a section of
\begin{equation}
 E\simeq\OO_{\PP^2}(4)\oplus\OO_{\PP^2}(2),\qquad
 \int_Sc_2(E)=8.
 \label{eq:exampleA-reflection-singularity}
\end{equation}
Its nonzero top Chern number forces a singularity on $S$ in every member. The equation at $x_1=0$ is $x_5P_5=0$, so $D_1$ contains $S$ together with a residual surface. The fixed component $S$ has Euler characteristic $3$.

\paragraph{Two fixed surfaces and twelve isolated points.}
For the geometry $(6,12021,6)$, in the order $x_0,\ldots,x_9$, the weights are
\begin{equation}
 \mathbf W=\begin{pmatrix}
 7&1&2&0&2&2&0&0&0&0\\
 3&0&1&1&1&1&0&0&0&0\\
 2&0&1&0&0&0&1&0&0&0\\
 -1&0&0&0&-1&-1&0&1&0&0\\
 2&0&0&0&0&1&0&0&1&0\\
 2&0&0&0&1&0&0&0&0&1
 \end{pmatrix},\qquad
 \deg P=(14,7,4,-2,4,4),
 \label{eq:p12021-weights}
\end{equation}
with trivial $G$ and
\begin{equation}
\begin{aligned}
 \mathcal I_{\rm SR}=\langle\,
 &x_0x_1,x_2x_6,x_3x_7,x_5x_8,x_6x_8,
 x_6x_9,x_7x_8,x_7x_9,x_8x_9,\\
 &x_0x_4x_5,x_0x_4x_9,x_1x_2x_5,x_1x_2x_7,\\
 &x_2x_3x_4,x_2x_3x_5,x_3x_4x_5,x_3x_4x_9,
 x_4x_5x_6\,\rangle.
\end{aligned}
\label{eq:p12021-SR}
\end{equation}
The reflection
\begin{equation}
 \sigma:(x_2,x_5,x_7)\longmapsto(-x_2,-x_5,-x_7)
 \label{eq:p12021-reflection}
\end{equation}
retains 46 of the 90 anticanonical monomials in the even sector. The fixed equations give the surfaces
\begin{equation}
 S_{35}=\{x_3=x_5=0\},\qquad S_{38}=\{x_3=x_8=0\},
 \label{eq:p12021-surfaces}
\end{equation}
with compensators
\begin{equation}
 \lambda_{S_{35}}=(1,-1,1,-1,1,1),\qquad
 \lambda_{S_{38}}=(1,-1,1,-1,-1,1).
 \label{eq:p12021-lambda}
\end{equation}
The retained monomials give
\begin{equation}
 P_{\rm sym}=x_3P_3+x_5x_8P_{58},
 \label{eq:p12021-polynomial}
\end{equation}
so both surfaces lie in every even member. They are disjoint by the SR generator $x_5x_8$. Each has four torus fixed points in smooth affine charts, giving
\begin{equation}
 \chi(S_{35})=\chi(S_{38})=4.
 \label{eq:p12021-surface-euler}
\end{equation}
Using $(a,b)=(\widehat D_3,\widehat D_5)$ and $(\widehat D_3,\widehat D_8)$ in \eqref{eq:normal-derivative-degree} yields
\begin{equation}
 \int_{S_{35}}c_2(E)=4,\qquad \int_{S_{38}}c_2(E)=8.
 \label{eq:p12021-singularity}
\end{equation}
Every member therefore has singular points on both fixed surfaces.

Away from the surfaces, nine one-dimensional ambient fixed strata remain. Seven have nonconstant hypersurface restrictions, listed in table~\ref{tab:p12021-roots}. For instance, on $x_0=x_2=x_9=0$ with the other coordinates nonzero, use
\begin{equation}
 z=\frac{x_5x_7}{x_1^2x_3x_8}.
 \label{eq:p12021-curve-coordinate}
\end{equation}
After removing an invertible monomial factor, the restricted equation is
\begin{equation}
 b_0+b_1z+b_2z^2+b_3z^3=0.
 \label{eq:p12021-cubic}
\end{equation}
The four coefficients are independent. For general coefficients, $b_0b_3\neq0$ and the discriminant is nonzero, giving three points in $\CC^*$. The remaining rows follow by the same method.

\begin{table}[!htbp]
\centering\small
\begin{tabular}{@{}lcc@{}}
\toprule
Vanishing coordinates & Polynomial degree & Points\\
\midrule
$\{x_0,x_2,x_4\}$ & 2 & 2\\
$\{x_0,x_2,x_9\}$ & 3 & 3\\
$\{x_0,x_4,x_6\}$ & 1 & 1\\
$\{x_1,x_2,x_9\}$ & 1 & 1\\
$\{x_1,x_4,x_6\}$ & 1 & 1\\
$\{x_2,x_5,x_7\}$ & 2 & 2\\
$\{x_5,x_6,x_7\}$ & 2 & 2\\
\bottomrule
\end{tabular}
\caption{Isolated fixed points of a general member of the singular
family $(6,12021,6)$. The polynomial degrees are measured in
quotient coordinates.}
\label{tab:p12021-roots}
\end{table}

The seven rows give $2+3+1+1+1+2+2=12$ points. The restrictions on the curves with zero sets $\{x_1,x_2,x_4\}$ and $\{x_2,x_4,x_9\}$ are nonzero multiples of $x_0^2x_3$, which is invertible there. This monomial also excludes the isolated ambient fixed points with zero sets $\{x_1,x_4,x_5,x_7\}$ and $\{x_1,x_4,x_5,x_9\}$. Thus the reduced fixed locus comprises two surfaces and twelve isolated points, with
\begin{equation}
 \chi(\Fix\sigma)=4+4+12=20.
 \label{eq:compensated-total}
\end{equation}
The total fixed-locus Euler characteristic remains $\chi(\Fix\sigma)=20$, as reported in the database~\cite{Cao:2024oqx}. However, the normal-derivative obstruction establishes that the entire even invariant family is singular, revising the original classification of this entry as a smooth orientifold.

\subsection{Involutions within and across toric phases}
\label{sec:involutions-across-triangulations}

We examine the proper NID exchanges across the sixteen FRSTs of polytope 431 at $h^{1,1}=4$. Three distinct coordinate involutions are supported by the same triangulation $T_0$, and each is compatible with four FRSTs. Together, they give twelve FRST--involution pairs across ten supporting triangulations. We write $T_i=(4,431,i)$ for the catalogue indices $i=0,\ldots,15$~\cite{Cao:2024oqx}.

In the order $x_0,\ldots,x_7$, the rays and weights are
\begin{equation}
 \mathsf V=\begin{pmatrix}
 1&0&0&0&-2&-1&-1&-1\\
 0&0&0&1&-1&-1&0&0\\
 0&0&1&0&-1&0&-1&0\\
 0&1&0&0&-1&0&0&-1
 \end{pmatrix},\qquad
 \mathbf W=\begin{pmatrix}
 2&1&1&1&1&0&0&0\\
 1&0&0&1&0&1&0&0\\
 1&0&1&0&0&0&1&0\\
 1&1&0&0&0&0&0&1
 \end{pmatrix}.
 \label{eq:poly431-matrices}
\end{equation}
The first four rays form a lattice basis, so $G$ is trivial. The three proper NID exchanges supported by $T_0$ are
\begin{equation}
\begin{aligned}
\sigma^{(1)}&:\quad x_1\longleftrightarrow x_2,\quad x_6\longleftrightarrow x_7,\\
\sigma^{(2)}&:\quad x_1\longleftrightarrow x_3,\quad x_5\longleftrightarrow x_7,\\
\sigma^{(3)}&:\quad x_2\longleftrightarrow x_3,\quad x_5\longleftrightarrow x_6.
\end{aligned}
\label{eq:poly431-exchanges}
\end{equation}
with zero-based coordinate indices. The two coordinates in each pair have distinct GLSM weights. On every supporting FRST listed in table~\ref{tab:poly431-support}, the exchange preserves the SR ideal, and its lattice lift preserves the fan by the correspondence of section~\ref{sec:slt-correspondence}.

Using the notation of~\eqref{eq:phase-action-relation},  where $\mathcal S(T)$ is the set of proper NID coordinate actions supported by $T$ and  $\mathcal T(\sigma)=\{T:\sigma\in\mathcal S(T)\}$,
Table~\ref{tab:poly431-support} gives the incidence relation for all sixteen FRSTs.
\begin{table}[!ht]
\centering\small
\setlength{\tabcolsep}{4pt}
\begin{tabular}{@{}l*{16}{c}@{}}
\toprule
Triangulation $i$ & $0$ & $1$ & $2$ & $3$ & $4$ & $5$ & $6$ & $7$
  & $8$ & $9$ & $10$ & $11$ & $12$ & $13$ & $14$ & $15$\\
\midrule
$\sigma^{(1)}$ & $\checkmark$ &  &  & $\checkmark$ &  &  &  &
  &  &  & $\checkmark$ &  &  &  &  & $\checkmark$\\
$\sigma^{(2)}$ & $\checkmark$ &  & $\checkmark$ &  &  &  &  &
  &  &  &  &  & $\checkmark$ & $\checkmark$ &  & \\
$\sigma^{(3)}$ & $\checkmark$ & $\checkmark$ &  &  &  &  &  &
  &  &  &  & $\checkmark$ &  &  & $\checkmark$ & \\
\bottomrule
\end{tabular}
\caption{Proper NID exchanges across the sixteen FRSTs of
polytope 431, giving twelve compatible FRST--involution
pairs on ten supporting triangulations.}
\label{tab:poly431-support}
\end{table}

Reading by triangulation gives $\mathcal S(T_0)=\{\sigma^{(1)},\sigma^{(2)},\sigma^{(3)}\}$, whereas $\mathcal S(T_{3,10,15})=\{\sigma^{(1)}\}$, $\mathcal S(T_{2,12,13})=\{\sigma^{(2)}\}$ and $\mathcal S(T_{1,11,14})=\{\sigma^{(3)}\}$. Conversely, the supporting triangulations for each involution are
\[
\begin{aligned}
 \mathcal T(\sigma^{(1)})&=\{T_0,T_3,T_{10},T_{15}\},\\
 \mathcal T(\sigma^{(2)})&=\{T_0,T_2,T_{12},T_{13}\},\\
 \mathcal T(\sigma^{(3)})&=\{T_0,T_1,T_{11},T_{14}\}.
\end{aligned}
\]
Thus $T_0$ is the unique common supporting fan for any two of the three actions, and
\[
 \sum_{i=0}^{15}|\mathcal S(T_i)|
 =\sum_{a=1}^{3}|\mathcal T(\sigma^{(a)})|=12.
\]

This example illustrates the many-to-many relation between toric phases and their compatible involutions. The phase-resolved data identify multiple orientifold choices on a single FRST, track individual involutions across different triangulations, and determine their common supporting phases. Together with the coordinate--lattice correspondence, they provide a basis for comparing invariant hypersurface families and orientifold geometry across toric phases.

\section{Catalogue update}
\label{sec:catalogue}
\label{sec:published-constructions}

We apply the fixed-locus calculations illustrated in section~\ref{sec:examples} to reassess and update the catalogue of~\cite{Cao:2024oqx}. Each orientifold record links a triangulation and a compatible involution to the chosen hypersurface eigenspace.  Table~\ref{tab:published-catalogues} summarises the scope of earlier constructions.

\begin{table}[!htbp]
\centering\small
\setlength{\tabcolsep}{4pt}
\begin{tabular}{@{}
>{\raggedright\arraybackslash}p{0.43\textwidth}
>{\raggedright\arraybackslash}p{0.41\textwidth}
c@{}}
\toprule
Construction & Scope & References\\
\midrule
Divisor exchanges and odd divisor classes
& $h^{1,1}\leq 4$
& \cite{Gao:2013pra}\\
\addlinespace
Divisor exchanges, fixed loci and Hodge splitting
& Catalogue through $h^{1,1}=6$
& \cite{Altman:2021pyc}\\
\addlinespace
Single-coordinate reflections; local and Whitney-brane D3 charges
& All FRSTs for $h^{1,1}\leq 7$; sampled triangulations for
  $8\leq h^{1,1}\leq 12$
& \cite{Crino:2022zjk}\\
\addlinespace
Divisor exchanges and multiple-coordinate reflections
& All FRSTs for $h^{1,1}\leq 7$; sampled triangulations for
  $8\leq h^{1,1}\leq 12$
& \cite{Cao:2024oqx}\\
\addlinespace
Lattice involutions, torus phases and invariant subdivisions
& General method; examples at $h^{1,1}=491$ and $243$
& \cite{Moritz:2023jdb}\\
\addlinespace
Ambient involutions and D3 charges
& CICYs and their conifold resolutions
& \cite{Carta:2020ohw}\\
\bottomrule
\end{tabular}
\caption{Earlier Calabi--Yau orientifold constructions.
The FRST ranges specify triangulation coverage for favourable
polytopes; the reflection sampling
of~\cite{Cao:2024oqx} is described below.
CICY denotes a complete-intersection Calabi--Yau threefold.}
\label{tab:published-catalogues}
\end{table}

The reflection scan in~\cite{Cao:2024oqx} includes all single, double and triple coordinate reflections through $h^{1,1}=6$. At $h^{1,1}=7$, it includes all single-coordinate reflections and samples 15 double and 15 triple reflections per triangulation. For $8\leq h^{1,1}\leq12$, it includes all single-coordinate reflections and samples 15 triple reflections per sampled triangulation. The update groups these coordinate sign patterns by their actions on the Cox quotient and adjusts the polynomial parity according to~\eqref{eq:reflection-eigenvalue}. For each triangulation, involution and hypersurface eigenspace, it records the defining equations, reduced components and Euler characteristics of the fixed locus.

We have recalculated 106 divisor-exchange records selected for previously identified fixed-locus discrepancies. Separate smoothness checks establish that each of the specified invariant families admits a smooth member. In 69 records, the isolated-point count changes while the sum of the fixed-surface Euler characteristics remains unchanged. In the remaining 37 records, the point count is unchanged and the surface Euler sum changes. Together, the updated records contain 112 fixed surfaces. The recalculated point counts, surface Euler characteristics and cubic self-intersections satisfy the holomorphic Lefschetz identity~\eqref{eq:index48} in all 106 cases. 

The ongoing  scan targets $17{,}414{,}478$ inputs from the divisor-exchange and reflection catalogues at $2\leq h^{1,1}\leq6$. An input specifies a polytope, a triangulation, an original catalogue record, a coordinate representative of the involution and a polynomial parity. At the calculation snapshot reported here, $9{,}377{,}208$ inputs have been processed. Of these, $256{,}381$ have a zero anticanonical eigenspace for the specified coordinate representative and parity, while $9{,}120{,}827$ admit nonzero invariant sections.

Fixed-locus data and smooth-model orientifold data are consequently distinguished in the update. The reduced fixed components of a singular invariant family can still be recorded, but the standard Hodge-splitting and D3-tadpole formulas are applied only when their smoothness and orientifold-sector assumptions have been established. The complete update will be made available through the \href{https://github.com/GroupofXG/anewcydatabase/}{GitHub}. The record structure is described in appendix~\ref{app:database-format}.

\section{Conclusions and outlook}
\label{sec:discussion}

We have reviewed the construction of Type IIB orientifolds on toric Calabi--Yau threefolds through divisor exchanges and coordinate reflections, relating the homogeneous-coordinate and lattice descriptions. The full Cox group, including finite identifications, determines the quotient actions and their fixed loci on invariant hypersurfaces. Together with adjunction and Lefschetz formulas, these constructions provide the geometric data for orientifold Hodge splittings and O-plane contributions to tadpole cancellation.

We have supplemented this framework with explicit calculations of fixed loci and orientifold charges. In Example A, the involution gives five isolated O3-planes and an O7 surface with Euler characteristic $27$. The corresponding D3 curvature budget for the specified local D7 configuration is $L_{\rm D3}=8$, and the complex-structure splitting is $(h^{2,1}_+,h^{2,1}_-)=(13,24)$. Finite Cox identifications give eight additional isolated O3-planes in a separate example. An important further result concerns the smoothness of invariant hypersurface families. The normal-derivative obstruction shows that certain reflection eigenspaces contain only singular hypersurfaces. In particular, the geometry $(6,12021,6)$ retains the fixed-locus Euler characteristic $\chi(\Fix\sigma)=20$, while its invariant family is necessarily singular. These calculations distinguish the topology of reduced fixed loci from the Hodge and charge data associated with smooth orientifolds.

An immediate goal is to complete the catalogue update, retaining the hypersurface family and verification status alongside each phase-action record. Our triangulation-wise enumeration identifies compatible involutions within and across toric phases. For polytope 431, three labelled NID exchanges give twelve compatible FRST--involution pairs across ten of the sixteen triangulations, with all three actions supported on $T_0$. The catalogue update also includes 106 recalculated divisor-exchange records whose fixed-locus data satisfy the holomorphic Lefschetz identity.
The next step is to complete the catalogue update, retaining
the hypersurface eigenspace and smoothness status alongside
each phase--action record.

Further applications include comparing local D7 stacks with recombined configurations such as Whitney branes~\cite{Crino:2022zjk,Collinucci:2008pf}, including their curvature and gauge-flux contributions. Combining these data with flux quantisation, tadpole cancellation and instanton zero-mode conditions provides geometric input for constructing compatible brane, flux and moduli-stabilisation sectors on the same compact threefold.

\section*{Acknowledgements}
XG thanks the organisers of the 2026 Calabi-Yau 50 conference for the invitation to contribute to these proceeding, and his collaborators for their contributions to the work reviewed here. We thank Yunhao Fu, Hancheng Li and Zhiwei Li for helpful discussions. XC and XG are supported by the National Natural Science Foundation of China (Grant No.~12375065) and the Natural Science Foundation of Sichuan Province (Grant No.~2026NSFSC0033). AI-assisted tools were used in preparing the manuscript; the authors take full responsibility for its content.

\appendix

\section{Database format}
\label{app:database-format}

\subsection{Database fields}

For each $h^{1,1}$, the catalogue has four JSON Lines files for polytopes, triangulations, involutions and orientifolds. Coordinate indices start at zero and follow the order of \texttt{rays}.

\begingroup \small
\medskip
\noindent\begin{minipage}{\linewidth}
\textbf{\texttt{polytopes.jsonl}}\par
\begin{itemize}
\setlength{\itemsep}{0pt}
\setlength{\parsep}{0pt}
\raggedright
\item \textbf{\texttt{polytope\_id}}: Polytope identifier.
\item \textbf{\texttt{legacy\_id}}: Polytope index in the original catalogue.
\item \textbf{\texttt{h11}}, \textbf{\texttt{h21}}: Hodge numbers $h^{1,1}(X)$ and $h^{2,1}(X)$.
\item \textbf{\texttt{rays}}: Ordered primitive rays $n_i$.
\item \textbf{\texttt{weights}}: GLSM weight matrix $\mathbf W$, with columns in ray order.
\item \textbf{\texttt{finite\_actions}}: Orders $d$ and exponents $a_i$ for generators of $G$, acting by $x_i\mapsto e^{2\pi i a_i/d}x_i$. The list is empty for trivial $G$.
\end{itemize}
\end{minipage}\par\medskip

\noindent\begin{minipage}{\linewidth}
\textbf{\texttt{triangulations.jsonl}}\par
\begin{itemize}
\setlength{\itemsep}{0pt}
\setlength{\parsep}{0pt}
\raggedright
\item \textbf{\texttt{triangulation\_id}}: FRST identifier.
\item \textbf{\texttt{polytope\_id}}: Link to the polytope record.
\item \textbf{\texttt{SR}}: Minimal Stanley--Reisner generators as lists of coordinate indices.
\item \textbf{\texttt{divisor\_basis}}: Basis classes $J_a$ expressed in the $D_i$.
\item \textbf{\texttt{divisor\_classes}}: Classes of the $D_i$ expressed in the $J_a$.
\item \textbf{\texttt{triple\_intersections}}: Nonzero $\int_XJ_aJ_bJ_c$ as \texttt{[[a,b,c], value]}, with $a\leq b\leq c$.
\item \textbf{\texttt{ambient\_intersections}}: Ambient quadruple intersections.
\item \textbf{\texttt{divisor\_hodge}}: Hodge numbers of the $D_i$ in ray order.
\end{itemize}
\end{minipage}\par\medskip

\noindent\begin{minipage}{\linewidth}
\textbf{\texttt{involutions.jsonl}}\par
\begin{itemize}
\setlength{\itemsep}{0pt}
\setlength{\parsep}{0pt}
\raggedright
\item \textbf{\texttt{involution\_id}}: Involution identifier.
\item \textbf{\texttt{polytope\_id}}: Link to the polytope record.
\item \textbf{\texttt{type}}: \texttt{"divisor\_exchange"} or \texttt{"reflection"}.
\item \textbf{\texttt{exchanged\_pairs}}: Pairs of exchanged coordinate indices.
\item \textbf{\texttt{flipped\_coordinates}}: Indices of coordinates whose signs are reversed.
\item \textbf{\texttt{reflection\_class}}: Sum of the rays for the flipped coordinates, taken in $N/2N$. Each nontrivial class realised by coordinate reflections has one representative per polytope.
\end{itemize}
\end{minipage}\par\medskip

\noindent\begin{minipage}{\linewidth}
\textbf{\texttt{orientifolds.jsonl}}\par
\begin{itemize}
\setlength{\itemsep}{0pt}
\setlength{\parsep}{0pt}
\raggedright
\item \textbf{\texttt{orientifold\_id}}: Orientifold identifier.
\item \textbf{\texttt{triangulation\_id}}: Link to the triangulation record.
\item \textbf{\texttt{involution\_id}}: Link to the involution record.
\item \textbf{\texttt{polynomial\_parity}}: Parity $\gamma$ of $P_{\rm sym}$.
\item \textbf{\texttt{omega\_parity}}: Sign of $\Omega_3$ under the involution.
\item \textbf{\texttt{polynomials}}: Named polynomials as lists of \texttt{[coefficient, exponent\_vector]} terms, with exponents in ray order.
\item \textbf{\texttt{smoothness}}: Smoothness status of a general member
of the specified invariant eigenspace, recorded as
\texttt{"smooth"}, \texttt{"singular"} or \texttt{null}
when unresolved.
\item \textbf{\texttt{hodge\_splitting}}: Odd Hodge dimensions
\texttt{h11\_minus} and \texttt{h21\_minus} for a verified
smooth member of the specified family. This field is
\texttt{null} when the family is known to be singular or
its smoothness has not been established. For a smooth member,
the even dimensions follow by subtraction from the Hodge
numbers of the corresponding smooth anticanonical family.
\item \textbf{\texttt{euler\_fixed}}: Euler characteristic $\chi(\Fix\sigma)$ of the full fixed locus.
\item \begin{minipage}[t]{\linewidth}
\textbf{\texttt{fixed\_loci}}: Fixed loci on $X$, with fields
  \begin{itemize}
  \setlength{\itemsep}{0pt}
  \setlength{\parsep}{0pt}
  \item \textbf{\texttt{label}}: Label within the record.
  \item \textbf{\texttt{type}}: O-plane type, or \texttt{null} if
unassigned.
  \item \textbf{\texttt{zero\_coordinates}}: Indices of coordinates set to zero.
  \item \textbf{\texttt{equations}}: Names of polynomials set to zero.
  \item \textbf{\texttt{euler}}: Euler characteristic of a positive-dimensional locus.
  \item \textbf{\texttt{point\_count}}: Number of distinct points in a zero-dimensional locus.
  \end{itemize}
\end{minipage}
\end{itemize}

\end{minipage}\par
\endgroup

\subsection{Records for Example A}

\begingroup
\small
\linespread{1}\selectfont
\setlength{\parindent}{0pt}
\setlength{\parskip}{3pt}
\raggedright

\lstset{
  style=databasejson, basicstyle=\ttfamily\footnotesize, columns=fullflexible, keepspaces=true, breaklines=true, breakatwhitespace=false, aboveskip=2pt, belowskip=3pt, numbers=none
}

These records describe Example A in section~\ref{sec:examplea}, with coordinate indices $0,\ldots,10$. Line breaks are added for display; each stored JSONL record occupies one line.

\textbf{\texttt{polytopes.jsonl}}\par\nobreak
\begin{lstlisting}
{"polytope_id":"P7_545", "legacy_id":545, "h11":7, "h21":37,
 "rays":[
   [1,0,0,0], [1,2,0,0], [1,0,2,0], [1,0,0,2],
   [-2,-1,-1,-1], [1,1,0,0], [1,0,1,0], [1,0,0,1],
   [1,1,1,0], [1,1,0,1], [1,0,1,1]],
 "weights":[
   [1,0,1,0,2,0,0,0,0,2,0], [0,1,1,0,0,0,0,0,-2,0,0],
   [0,0,-1,1,0,0,0,0,2,-2,0], [0,0,1,0,1,1,0,0,-1,1,0],
   [0,0,0,0,1,0,1,0,0,1,0], [0,0,0,0,1,0,0,1,1,0,0],
   [0,0,-1,0,0,0,0,0,1,-1,1]],
 "finite_actions":[]}
\end{lstlisting}

\textbf{\texttt{triangulations.jsonl}} (excerpt).
The SR generators are those of~\eqref{eq:exampleA-SR}.\par\nobreak
\begin{lstlisting}
{"triangulation_id":"P7_545_TA", "polytope_id":"P7_545",
 "SR":[[0,1],[0,2],[0,3],[0,8],[0,9],[0,10],
       [1,2],[1,3],[1,6],[1,7],[1,10],
       [2,3],[2,5],[2,7],[2,9],
       [5,10],[6,9],[7,8],[7,9],[7,10],[9,10],
       [3,4,5,6],[3,4,5,8],[3,4,6,8],
       [4,5,6,7],[4,5,8,9],[4,6,8,10]]}
\end{lstlisting}

\textbf{\texttt{involutions.jsonl}} (exchange and reflection).\par\nobreak
\begin{lstlisting}
{"involution_id":"P7_545_I1", "polytope_id":"P7_545",
 "type":"divisor_exchange", "exchanged_pairs":[[0,1],[6,8],[7,9]]}
{"involution_id":"P7_545_R1", "polytope_id":"P7_545",
 "type":"reflection", "flipped_coordinates":[0],
 "reflection_class":[1,0,0,0]}
\end{lstlisting}

\textbf{\texttt{orientifolds.jsonl}}.
\texttt{f\_minus} and \texttt{f\_plus} are $y_9$ and $y_8$ of~\eqref{eq:exampleA-generators}; locus labels follow table~\ref{tab:exampleA}.\par\nobreak
\begin{lstlisting}
{"orientifold_id":"P7_545_TA_I1_Pplus",
 "triangulation_id":"P7_545_TA", "involution_id":"P7_545_I1",
 "polynomial_parity":1, "omega_parity":-1, "smoothness":"smooth",
 "polynomials":{
   "f_minus":[[1,[2,0,0,0,0,0,1,1,0,0,0]],
              [-1,[0,2,0,0,0,0,0,0,1,1,0]]],
   "f_plus":[[1,[2,0,0,0,0,0,1,1,0,0,0]],
             [1,[0,2,0,0,0,0,0,0,1,1,0]]]},
 "fixed_loci":[
   {"label":1, "type":"O7", "zero_coordinates":[],
    "equations":["f_minus"], "euler":27},
   {"label":2, "type":"O3", "zero_coordinates":[3,4,10],
    "equations":["f_plus"], "point_count":1},
   {"label":3, "type":"O3", "zero_coordinates":[2,4,10],
    "equations":["f_plus"], "point_count":1},
   {"label":4, "type":"O3", "zero_coordinates":[3,5],
    "equations":["f_plus"], "point_count":3}],
 "hodge_splitting":{"h11_minus":2, "h21_minus":24},
 "euler_fixed":32}
\end{lstlisting}

\par\endgroup

\bibliographystyle{JHEP}
\bibliography{refree}

\providecommand{\href}[2]{#2}\begingroup\raggedright\begin{thebibliography}{10}

\bibitem{Calabi1957}
E.~Calabi, \emph{On {K\"ahler} manifolds with vanishing canonical class},  in
  \emph{Algebraic Geometry and Topology: A Symposium in Honor of Solomon
  Lefschetz}, R.H.~Fox, D.C.~Spencer and A.W.~Tucker, eds., (Princeton, NJ),
  pp.~78--89, Princeton University Press (1957).

\bibitem{Yau:1977ms}
S.-T.~Yau, \emph{{Calabi}'s conjecture and some new results in algebraic
  geometry}, \href{https://doi.org/10.1073/pnas.74.5.1798}{\emph{Proc. Natl.
  Acad. Sci. U.S.A.} {\bfseries 74} (1977) 1798}.

\bibitem{Yau:1978cfy}
S.-T.~Yau, \emph{On the {Ricci} curvature of a compact {K\"ahler} manifold and
  the complex {Monge--Amp\`ere} equation, {I}},
  \href{https://doi.org/10.1002/cpa.3160310304}{\emph{Commun. Pure Appl. Math.}
  {\bfseries 31} (1978) 339}.

\bibitem{Candelas:1985en}
P.~Candelas, G.T.~Horowitz, A.~Strominger and E.~Witten, \emph{Vacuum
  configurations for superstrings},
  \href{https://doi.org/10.1016/0550-3213(85)90602-9}{\emph{Nucl. Phys. B}
  {\bfseries 258} (1985) 46}.

\bibitem{Candelas:1990rm}
P.~Candelas, X.C.~de~la Ossa, P.S.~Green and L.~Parkes, \emph{A pair of
  {Calabi--Yau} manifolds as an exactly soluble superconformal theory},
  \href{https://doi.org/10.1016/0550-3213(91)90292-6}{\emph{Nucl. Phys. B}
  {\bfseries 359} (1991) 21}.

\bibitem{Bershadsky:1993ta}
M.~Bershadsky, S.~Cecotti, H.~Ooguri and C.~Vafa, \emph{Holomorphic anomalies
  in topological field theories},
  \href{https://doi.org/10.1016/0550-3213(93)90548-4}{\emph{Nucl. Phys. B}
  {\bfseries 405} (1993) 279}
  [\href{https://arxiv.org/abs/hep-th/9302103}{{\ttfamily hep-th/9302103}}].

\bibitem{Bershadsky:1993cx}
M.~Bershadsky, S.~Cecotti, H.~Ooguri and C.~Vafa, \emph{{Kodaira--Spencer}
  theory of gravity and exact results for quantum string amplitudes},
  \href{https://doi.org/10.1007/BF02099774}{\emph{Commun. Math. Phys.}
  {\bfseries 165} (1994) 311}
  [\href{https://arxiv.org/abs/hep-th/9309140}{{\ttfamily hep-th/9309140}}].

\bibitem{Strominger:1996it}
A.~Strominger, S.-T.~Yau and E.~Zaslow, \emph{Mirror symmetry is {T}-duality},
  \href{https://doi.org/10.1016/0550-3213(96)00434-8}{\emph{Nucl. Phys. B}
  {\bfseries 479} (1996) 243}
  [\href{https://arxiv.org/abs/hep-th/9606040}{{\ttfamily hep-th/9606040}}].

\bibitem{Sagnotti:1987tw}
A.~Sagnotti, \emph{{Open Strings and their Symmetry Groups}},  in \emph{{NATO
  Advanced Summer Institute on Nonperturbative Quantum Field Theory (Cargese
  Summer Institute)}}, 9, 1987
  [\href{https://arxiv.org/abs/hep-th/0208020}{{\ttfamily hep-th/0208020}}].

\bibitem{Brunner:2003zm}
I.~Brunner and K.~Hori, \emph{Orientifolds and mirror symmetry},
  \href{https://doi.org/10.1088/1126-6708/2004/11/005}{\emph{JHEP} {\bfseries
  11} (2004) 005} [\href{https://arxiv.org/abs/hep-th/0303135}{{\ttfamily
  hep-th/0303135}}].

\bibitem{Grimm:2004uq}
T.W.~Grimm and J.~Louis, \emph{{The effective action of $\mathcal{N}=1$
  Calabi--Yau orientifolds}},
  \href{https://doi.org/10.1016/j.nuclphysb.2004.08.005}{\emph{Nucl. Phys. B}
  {\bfseries 699} (2004) 387}
  [\href{https://arxiv.org/abs/hep-th/0403067}{{\ttfamily hep-th/0403067}}].

\bibitem{Jockers:2004yj}
H.~Jockers and J.~Louis, \emph{{The effective action of D7-branes in
  $\mathcal{N}=1$ Calabi--Yau orientifolds}},
  \href{https://doi.org/10.1016/j.nuclphysb.2004.11.009}{\emph{Nucl. Phys. B}
  {\bfseries 705} (2005) 167}
  [\href{https://arxiv.org/abs/hep-th/0409098}{{\ttfamily hep-th/0409098}}].

\bibitem{Blumenhagen:2008zz}
R.~Blumenhagen, V.~Braun, T.W.~Grimm and T.~Weigand, \emph{{GUTs in Type IIB
  orientifold compactifications}},
  \href{https://doi.org/10.1016/j.nuclphysb.2009.02.011}{\emph{Nucl. Phys. B}
  {\bfseries 815} (2009) 1} [\href{https://arxiv.org/abs/0811.2936}{{\ttfamily
  0811.2936}}].

\bibitem{Giddings:2001yu}
S.B.~Giddings, S.~Kachru and J.~Polchinski, \emph{{Hierarchies from fluxes in
  string compactifications}},
  \href{https://doi.org/10.1103/PhysRevD.66.106006}{\emph{Phys. Rev. D}
  {\bfseries 66} (2002) 106006}
  [\href{https://arxiv.org/abs/hep-th/0105097}{{\ttfamily hep-th/0105097}}].

\bibitem{Gukov:1999ya}
S.~Gukov, C.~Vafa and E.~Witten, \emph{{CFT's from Calabi--Yau four-folds}},
  \href{https://doi.org/10.1016/S0550-3213(00)00373-4}{\emph{Nucl. Phys. B}
  {\bfseries 584} (2000) 69}
  [\href{https://arxiv.org/abs/hep-th/9906070}{{\ttfamily hep-th/9906070}}].

\bibitem{Kachru:2003aw}
S.~Kachru, R.~Kallosh, A.D.~Linde and S.P.~Trivedi, \emph{{De Sitter vacua in
  string theory}},
  \href{https://doi.org/10.1103/PhysRevD.68.046005}{\emph{Phys. Rev. D}
  {\bfseries 68} (2003) 046005}
  [\href{https://arxiv.org/abs/hep-th/0301240}{{\ttfamily hep-th/0301240}}].

\bibitem{Balasubramanian:2005zx}
V.~Balasubramanian, P.~Berglund, J.P.~Conlon and F.~Quevedo, \emph{{Systematics
  of moduli stabilisation in Calabi--Yau flux compactifications}},
  \href{https://doi.org/10.1088/1126-6708/2005/03/007}{\emph{JHEP} {\bfseries
  03} (2005) 007} [\href{https://arxiv.org/abs/hep-th/0502058}{{\ttfamily
  hep-th/0502058}}].

\bibitem{Batyrev:1993oya}
V.V.~Batyrev, \emph{{Dual polyhedra and mirror symmetry for Calabi--Yau
  hypersurfaces in toric varieties}}, {\emph{J. Algebraic Geom.} {\bfseries 3}
  (1994) 493} [\href{https://arxiv.org/abs/alg-geom/9310003}{{\ttfamily
  alg-geom/9310003}}].

\bibitem{Cox:1993fz}
D.A.~Cox, \emph{The homogeneous coordinate ring of a toric variety}, {\emph{J.
  Algebraic Geom.} {\bfseries 4} (1995) 17}
  [\href{https://arxiv.org/abs/alg-geom/9210008}{{\ttfamily
  alg-geom/9210008}}].

\bibitem{CoxLittleSchenck}
D.A.~Cox, J.B.~Little and H.K.~Schenck, \emph{{Toric Varieties}}, vol.~124 of
  \emph{Graduate Studies in Mathematics}, American Mathematical Society (2011),
  \href{https://doi.org/10.1090/gsm/124}{10.1090/gsm/124}.

\bibitem{Kreuzer:2000xy}
M.~Kreuzer and H.~Skarke, \emph{{Complete classification of reflexive polyhedra
  in four dimensions}},
  \href{https://doi.org/10.4310/ATMP.2000.v4.n6.a2}{\emph{Adv. Theor. Math.
  Phys.} {\bfseries 4} (2000) 1209}
  [\href{https://arxiv.org/abs/hep-th/0002240}{{\ttfamily hep-th/0002240}}].

\bibitem{Altman:2014bfa}
R.~Altman, J.~Gray, Y.-H.~He, V.~Jejjala and B.D.~Nelson, \emph{{A Calabi--Yau
  database: threefolds constructed from the Kreuzer--Skarke list}},
  \href{https://doi.org/10.1007/JHEP02(2015)158}{\emph{JHEP} {\bfseries 02}
  (2015) 158} [\href{https://arxiv.org/abs/1411.1418}{{\ttfamily 1411.1418}}].

\bibitem{Demirtas:2022hqf}
M.~Demirtas, A.~Rios-Tascon and L.~McAllister, \emph{{CYTools: a software
  package for analyzing Calabi--Yau manifolds}},
  \href{https://arxiv.org/abs/2211.03823}{{\ttfamily 2211.03823}}.

\bibitem{Witten:1993yc}
E.~Witten, \emph{{Phases of $N=2$ theories in two dimensions}},
  \href{https://doi.org/10.1016/0550-3213(93)90033-L}{\emph{Nucl. Phys. B}
  {\bfseries 403} (1993) 159}
  [\href{https://arxiv.org/abs/hep-th/9301042}{{\ttfamily hep-th/9301042}}].

\bibitem{Gao:2013pra}
X.~Gao and P.~Shukla, \emph{{On classifying the divisor involutions in
  Calabi--Yau threefolds}},
  \href{https://doi.org/10.1007/JHEP11(2013)170}{\emph{JHEP} {\bfseries 11}
  (2013) 170} [\href{https://arxiv.org/abs/1307.1139}{{\ttfamily 1307.1139}}].

\bibitem{Altman:2021pyc}
R.~Altman, J.~Carifio, X.~Gao and B.D.~Nelson, \emph{{Orientifold Calabi--Yau
  threefolds with divisor involutions and string landscape}},
  \href{https://doi.org/10.1007/JHEP03(2022)087}{\emph{JHEP} {\bfseries 03}
  (2022) 087} [\href{https://arxiv.org/abs/2111.03078}{{\ttfamily
  2111.03078}}].

\bibitem{Crino:2022zjk}
C.~Crin{\`o}, F.~Quevedo, A.~Schachner and R.~Valandro, \emph{{A database of
  Calabi--Yau orientifolds and the size of D3-tadpoles}},
  \href{https://doi.org/10.1007/JHEP08(2022)050}{\emph{JHEP} {\bfseries 08}
  (2022) 050} [\href{https://arxiv.org/abs/2204.13115}{{\ttfamily
  2204.13115}}].

\bibitem{Cao:2024oqx}
X.~Cao, H.~Gao and X.~Gao, \emph{{Orientifold Calabi--Yau threefolds: divisor
  exchanges and multi-reflections}},
  \href{https://doi.org/10.1007/JHEP10(2024)188}{\emph{JHEP} {\bfseries 10}
  (2024) 188} [\href{https://arxiv.org/abs/2407.02565}{{\ttfamily
  2407.02565}}].

\bibitem{Carta:2020ohw}
F.~Carta, J.~Moritz and A.~Westphal, \emph{{A landscape of orientifold vacua}},
  \href{https://doi.org/10.1007/JHEP05(2020)107}{\emph{JHEP} {\bfseries 05}
  (2020) 107} [\href{https://arxiv.org/abs/2003.04902}{{\ttfamily
  2003.04902}}].

\bibitem{Moritz:2023jdb}
J.~Moritz, \emph{{Orientifolding Kreuzer--Skarke}},
  \href{https://arxiv.org/abs/2305.06363}{{\ttfamily 2305.06363}}.

\bibitem{Batyrev:2005jc}
V.V.~Batyrev and M.~Kreuzer, \emph{{Integral cohomology and mirror symmetry for
  Calabi--Yau 3-folds}},  in \emph{{Mirror Symmetry V}}, vol.~38 of
  \emph{AMS/IP Studies in Advanced Mathematics}, pp.~255--270, American
  Mathematical Society (2006)
  [\href{https://arxiv.org/abs/math/0505432}{{\ttfamily math/0505432}}].

\bibitem{AtiyahBott1968}
M.F.~Atiyah and R.~Bott, \emph{{A Lefschetz fixed point formula for elliptic
  complexes: II. Applications}},
  \href{https://doi.org/10.2307/1970721}{\emph{Ann. Math.} {\bfseries 88}
  (1968) 451}.

\bibitem{Atiyah:1968rj}
M.F.~Atiyah and G.B.~Segal, \emph{{The index of elliptic operators: II}},
  \href{https://doi.org/10.2307/1970716}{\emph{Ann. Math.} {\bfseries 87}
  (1968) 531}.

\bibitem{Collinucci:2008pf}
A.~Collinucci, F.~Denef and M.~Esole, \emph{{D-brane deconstructions in IIB
  orientifolds}},
  \href{https://doi.org/10.1088/1126-6708/2009/02/005}{\emph{JHEP} {\bfseries
  02} (2009) 005} [\href{https://arxiv.org/abs/0805.1573}{{\ttfamily
  0805.1573}}].

\bibitem{Green:1996dd}
M.B.~Green, J.A.~Harvey and G.W.~Moore, \emph{{I-brane inflow and anomalous
  couplings on D-branes}},
  \href{https://doi.org/10.1088/0264-9381/14/1/008}{\emph{Class. Quant. Grav.}
  {\bfseries 14} (1997) 47}
  [\href{https://arxiv.org/abs/hep-th/9605033}{{\ttfamily hep-th/9605033}}].

\bibitem{Morales:1998ux}
J.F.~Morales, C.A.~Scrucca and M.~Serone, \emph{{Anomalous couplings for
  D-branes and O-planes}},
  \href{https://doi.org/10.1016/S0550-3213(99)00217-5}{\emph{Nucl. Phys. B}
  {\bfseries 552} (1999) 291}
  [\href{https://arxiv.org/abs/hep-th/9812071}{{\ttfamily hep-th/9812071}}].

\end{thebibliography}\endgroup

\end{document}